\documentclass[12pt]{iopart}

\usepackage[english]{babel}
\usepackage[utf8]{inputenc}
\usepackage{url}

\usepackage{iopams}
\expandafter\let\csname equation*\endcsname\relax

\expandafter\let\csname endequation*\endcsname\relax
\usepackage{amsmath}
\usepackage{bm}

\usepackage{color}
\usepackage{graphicx}
\usepackage{caption}

\usepackage[a4paper]{geometry}

\usepackage{siunitx}

\usepackage[version=4]{mhchem}

\usepackage{hyperref}
\hypersetup{pdftex,colorlinks=true,allcolors=black}
\usepackage{hypcap}

\newcommand {\brc}[1]{\left( #1 \right)}
\newcommand {\sqrbrc}[1]{\left[ #1 \right]}
\newcommand {\crlbrc}[1]{\left\{ #1 \right\}}
\newcommand {\avr}[1]{\left<{#1}\right>}
\newcommand {\txt}[1]{\textrm{#1}}
\newcommand {\abs}[1]{\left|{#1}\right|}

\renewcommand {\vec}[1]{\boldsymbol #1 }

\begin{document}

\title{Minkowski functionals for phase behavior under confinement}

\author{Arnout M P Boelens and Hamdi A Tchelepi}

\address{Department of Energy Resources Engineering, Stanford University, Stanford,  California  94305, USA}
\ead{tchelepi@stanford.edu}
\vspace{10pt}
\begin{indented}
\item[]\today
\end{indented}

\begin{abstract}
In this work, the Minkowski functionals are used as a framework to study how
morphology (i.e. the shape of a structure) and topology (i.e. how different
structures are connected) influence wall adsorption and capillary condensation under
tight confinement. Numerical simulations based on classical density functional theory (DFT) are run for a
wide variety of geometries using both hard-sphere and Lennard-Jones fluids.
These DFT computations are compared to results obtained using the Minkowski
functionals. It is found that the Minkowski functionals can provide a good
description of the behavior of Lennard-Jones fluids down to small system
sizes. In addition, through decomposition of the free energy, the Minkowski
functionals provide a good framework to better understand what are the dominant
contributions to the physics of a system. Lastly, while studying the phase
envelope shift as a function of the
Minkowski functionals it is found that topology has a different effect depending
on whether the phase transition under consideration is a first- or a second-order
transition.
\end{abstract}

\vspace{2pc}

\submitto{\JPCM}

\section{Introduction}

Under tight confinement a gas can form a condensed phase at a pressure below the
bulk vapor pressure. This phenomenon is known as capillary condensation and has
applications in many fields of science and engineering, including the storage of hydrogen carriers
\cite{berube2007,he2014,he2016}, battery technology \cite{li2017b}, hydrocarbons
extraction from unconventional reservoirs \cite{barsotti2016}, and carbon dioxide sequestration
\cite{belmabkhout2009}. Capillary condensation can have a large effect on
transport properties \cite{bui2016}, and it is reported in the literature that
both morphology (i.e. the shape of a structure) and topology (i.e. how different
structures are connected) have a strong effect on the sorption of both sub- and
supercritical fluids \cite{he2014,melnichenko2016,ghosh2018}. For ordered porous
media, the relation between capillary condensation and geometry is well
understood \cite{barrett1951}; however, in practice many porous media are
disordered rather than ordered. Although simple geometries like cylinders, slit
pores, ink bottles, and spheres \cite{douglas2003,coasne2007} have been studied
extensively, capillary condensation in disordered porous media is not well
understood \cite{mason1982,mason1988,sarkisov2001,libby2004,coasne2007,coasne2013}. 

In this work, we study capillary condensation and wall adsorption under
confinement (i.e. small pores) through the lens of the
Minkowski functionals. These functionals are a concept from integral geometry
which not only characterize the morphology, but also the topology of spatial
patterns \cite{mecke1998}, and they have been applied in a wide array of
research areas including astronomy \cite{banados1994,schmalzing1998},
statistical physics \cite{mecke1991}, phase behavior \cite{mecke1997}, and
granular materials \cite{scheel2008,saadatfar2017}. For a system in $D$
dimensions, there are $D+1$ Minkowski functionals and in the case of a
two-dimensional system these functionals are related to the surface area,
circumference, and signed curvature (i.e.\ the Euler characteristic) of the
system \cite{mecke2000}. In addition to providing a method to characterize
spatial patterns, the Minkowski functionals also provide a powerful connection
between the thermodynamics and the geometry of a system. In many cases, the free energy of a system can be
expressed as a linear combination of Minkowski functionals
\cite{hadwiger1957,konig2004}. Once an expression for the free energy has been
found, other thermodynamic properties can be derived including the surface
tension, excess adsorption, and shifts in the phase envelope \cite{mecke2005}.

Minkowski functionals can be used in combination with experiments, theory, or
simulations. In this work, we employ classical density functional
theory (DFT) \cite{evans1979} to compute the free energy and adsorption for a
wide variety of geometries for both hard-sphere and Lennard-Jones fluids. These
results are then compared to results obtained using the Minkowski functionals.
The Minkowski functionals have mainly been used to study hard-sphere
fluids \cite{konig2004}. In this study, we find that Minkowski
functionals can also provide a good description of Lennard-Jones fluids down to
fairly small system sizes. This, in turn, means that the decomposition of the
free energy given by the Minkowski functionals can provide valuable insight into
the physical behavior of a Lennard-Jones fluid under confinement; e.g. it is found that topology has a different effect on the phase envelope
shift of a Lennard-Jones fluid under confinement depending on whether the phase
transition under consideration is a first- or a second-order transition. It is
left for future research to investigate whether this behavior is specific to
wall adsorption and capillary condensation, or whether this is a more universal
phenomenon.

\section{Theory}

\subsection{Minkowski functionals}
\label{sec:minkowski}

The Minkowski functionals are a concept from integral geometry. These
functionals characterize both the morphology and the topology of a spatial pattern
\cite{minkowski1903,schneider2013}. 
For a $D$ dimensional space, there are $D+1$ functionals. 
Considering a 2D system with a surface, $X$, and a smooth boundary, $\delta X$,
the following functionals can be defined:
\begin{equation}
\arraycolsep=1.4pt\def\arraystretch{2.2}
\begin{array}{lll}
\displaystyle M_{0} \brc{X} = \int_{\delta X} d A                       & = A \brc{X}: & \txt{ Surface area,} \\
\displaystyle M_{1} \brc{X} = \frac{1}{2} \int_{\delta X} d L           & = C \brc{X}: & \txt{ Circumference,} \\
\displaystyle M_{2} \brc{X} = \frac{1}{2} \int_{\delta X} k \brc{X} d L & = K \brc{X}: & \txt{ Signed curvature ($= \pi \; \chi$),} 
\end{array} 
\end{equation}
where $d A$ is a surface element, $d L$ is a circumference element, and $k
\brc{X}$ is the signed curvature \cite{legland2011}. Following the Gauss-Bonnet
theorem, the signed curvature is directly proportional to the Euler
characteristic, $\chi$, which is a measure of connectivity/topology.
Now consider a functional, ${\mathcal M} \brc{X}$, which is additive:
\begin{equation}
  {\mathcal M} \brc{X_{1} \cup X_{2}}
=           
  {\mathcal M} \brc{X_{1}}
+ {\mathcal M} \brc{X_{2}}
- {\mathcal M} \brc{X_{1} \cap X_{2}},
\label{eqn:additive}
\end{equation}
motion invariant:
\begin{equation}
  {\mathcal M} \brc{g X}
=           
  {\mathcal M} \brc{X},
\end{equation}
and continuous:
\begin{equation}
  {\mathcal M} \brc{X_{n}}
\to           
  {\mathcal M} \brc{X}
\;\;\; \txt{\it if: } \;
  X_{n} \to X
\;\;\; \txt{\it for: } \;
  n \to \infty.
\end{equation}
Then, following Hadwiger's theorem \cite{hadwiger1957}, this functional,
${\mathcal M} \brc{X}$, can be expressed as a linear combination of Minkowski functionals, $M_{\nu} \brc{X}$, as follows:
\begin{equation}
  {\mathcal M} \brc{X}
=
  \sum_{\nu=0}^{d} c_{\nu} M_{\nu} \brc{X}           
\end{equation}
An example of such a functional is the grand potential or Landau free energy,
$\Omega \brc{\mu, V, T}$, which is commonly used to study capillary condensation
\cite{kierlik2001,ravikovitch2006}. For a 2D system, the grand potential can
thus be written as \cite{mecke1998}:
\begin{equation}
  \frac{\Omega \brc{X}}{L}
=
- p \brc{\mu, T}      A \brc{X}
+ \sigma \brc{\mu, T} C \brc{X}
+ \kappa \brc{\mu, T} K \brc{X},
\end{equation}
where $L$ is a unit length, $p \brc{\mu, T}$ is the pressure, $\sigma
\brc{\mu, T}$ is the surface tension, and $\kappa \brc{\mu, T}$ is the signed bending rigidity.
The above expression for the grand potential demonstrates the importance of
Hadwiger's theorem. The pressure, surface tension, and signed bending rigidity are all only
dependent on the chemical potential, $\mu$, and the temperature, $T$. Thus,
the above equation shows how thermodynamics can be separated from morphology and
topology \cite{konig2004}. Once the grand potential is known, other thermodynamic
properties can be derived. This includes the excess free energy or surface
tension \cite{douglas2003,mecke2005}:
\begin{align}
  \gamma
=&
  \frac{\brc{\Omega/L + p_{b} A}}{C} \\
=&
  \brc{p_{b} \brc{\mu, T} - p \brc{\mu, T}} \frac{A \brc{X}}{C \brc{X}}
+ \sigma \brc{\mu, T} 
+ \kappa \brc{\mu, T} \frac{K \brc{X}}{C \brc{X}},
\end{align}
and, through Gibbs' theorem, the excess adsorption:
\begin{align}
- \Gamma_{\txt{ex}}
=&
  \frac{1}{C} \int  \rho_{b} - \rho \brc{r} \; d A 
=
  \brc{\frac{\partial \gamma}{\partial \mu}}_{T,V} \\
=&
  \frac{\partial}{\partial \mu}
  \brc{p_{b} - p \brc{\mu, T}} \frac{A \brc{X}}{C \brc{X}}
+ \frac{\partial \sigma}{\partial \mu}
+ \frac{\partial \kappa}{\partial \mu} \frac{K \brc{X}}{C \brc{X}}. 
\label{eqn:excess}
\end{align}
In the above equations, $p_{b}$ is the bulk pressure and $\rho_{b}$
is the bulk density. While it has been suggested in the past that Gibbs' theorem
is not valid for some systems undergoing capillary condensation
\cite{kierlik2001,kierlik2002}, later work found that Gibbs' theorem is not
violated when using an arc length tracking algorithm
\cite{salinger2003,douglas2003}.

In addition to the excess adsorption, one can also compute the effect of
confinement on the phase envelope \cite{mecke2005}:
\begin{equation}
 \Delta p \brc{\mu, T}
=
  \sigma_{lg}' \brc{\mu, T} \frac{C \brc{X}}{A \brc{X}}
+ \kappa_{lg}' \brc{\mu, T} \frac{K \brc{X}}{A \brc{X}},
\label{eqn:phase}
\end{equation}
with:
\begin{align}
  \sigma_{lg}' \brc{\mu, T} =& \sigma_{sg} \brc{\mu, T} - \sigma_{sl} \brc{\mu, T},   \\
  \kappa_{lg}' \brc{\mu, T} =& \kappa_{sg} \brc{\mu, T} - \kappa_{sl} \brc{\mu, T}, 
\end{align}
where $\sigma_{sg} \brc{\mu, T}$ and $\sigma_{sl} \brc{\mu, T}$ are the solid-gas
and solid-liquid surface energies, respectively, and $\kappa_{sg} \brc{\mu, T}$
and $\kappa_{sl} \brc{\mu, T}$ are the solid-gas and solid-liquid bending
rigidities. The above equation is a generalization of the Kelvin equation, and an
equation of the same form can be derived for the temperature shift \cite{evans1987}.
Two points should be noted about the above equation: i) due to diverging density
fluctuations, a mean field approach is not expected to
fully capture the correct scaling at the critical point \cite{evans1990}, and ii) the correlation
length that measures the range of density fluctuations at the critical point
also diverges, resulting in a potential violation of Hadwiger's theorem.
However, while these points need to be investigated further, the above equation
should give a good first approximation of how phase behavior is affected by
topology. 

\section{Methods}

\subsection{Density Functional Theory}
The Minkowski functionals can be used with either experiments, theory, or
simulations. In this work, we decided on using classical density functional
theory (DFT) to compute the coefficients in front of the Minkowski functionals.
DFT is a mean field approach which was first developed for quantum mechanics
\cite{hohenberg1964}, but was later adapted to describe classical mechanical
systems \cite{evans1979} as well. This mean field approach has the advantage of
giving a description of the physics at the nanoscopic molecular level, while
scaling up to the mesoscopic level at which capillary condensation occurs.

The two basic assumptions of density functional theory are; i) the Hohenberg-Kohn
variational principle, which states that there is a functional of the ground
state free energy which can be fully recovered from the ground-state one-particle density
distribution, and ii) the Gibbs' inequality, which states that any particle
density distribution that is not the ground state will have a higher free
energy  than the ground state free energy \cite{hohenberg1964}. 
Formulated in the grand canonical ($\mu$,$V$,$T$) ensemble, at the most basic level
this means that classical DFT solves the following minimization problem:
\begin{equation}
  \frac{\delta \Omega}{\delta \rho \brc{\vec{r}}}
=
  0, 
\end{equation}
where $\Omega$ is the grand potential or Landau free energy, $\rho$ is the density, and
$\delta$ is the Fr\'{e}chet (functional) derivative \cite{heroux2007}. To solve
the above equation, we use the DFT solver Tramonto, which is
developed at the Sandia National Laboratories
\cite{frink2000a,frink2000b,salinger2003,douglas2003,heroux2007}. This code uses perturbation theory
where the grand potential is split up as:
\begin{equation}
  \Omega
=
  F_{\txt{id}}
+ F_{\txt{hs}}
+ F_{\txt{p}}
- \int d \vec{r} \rho \brc{\vec{r}} \sqrbrc{V \brc{\vec{r}} - \mu},
\end{equation}
where $F_{\txt{id}}$ is the ideal contribution, $F_{\txt{hs}}$ is the
hard-sphere contribution, and $F_{\txt{p}}$ is the perturbation contribution. $V
\brc{\vec{r}}$ is the external potential resulting from (pore) walls acting on
the fluid. The individual contributions are given by the following integrals:
\begin{align}
  F_{\txt{id}}
&=
  \beta^{-1} \int d \vec{r} \rho \brc{\vec{r}}
  \crlbrc{ \ln{\sqrbrc{ \Lambda^{3} \rho \brc{\vec{r}}  } - 1 } } \\
  F_{\txt{hs}}
&=
  \int d \vec{r} \Phi \crlbrc{ \bar{\rho}_{\gamma} \brc{\vec{r}} } \\
  F_{\txt{p}}
&=
  \frac{1}{2} \int d \vec{r} \int \vec{r}' \rho \brc{\vec{r}} \rho \brc{\vec{r}'} U_{p} \brc{|\vec{r} - \vec{r}' |}.
\end{align}
In the above equations, $\beta^{-1} = k_{B} T$, with $k_{B}$ the Boltzmann
constant, and $T$ the temperature, $\Lambda$ is the thermal de Broglie wavelength,
$\Phi$ is the excess free energy density which is a function of
$\bar{\rho}_{\gamma}$, a set of weighted non-local densities, and $U_{p}$ is an
interaction potential. This potential is based on the Weeks-Chandler-Anderson
approach \cite{weeks1971}, which splits an interaction potential as $U_{p}
\brc{r} = u \brc{r_{\txt{min}} }$ for $r < r_{\txt{min}}$ and $U_{p}
\brc{r} = u \brc{r}$ for $r \geq r_{\txt{min}}$. The potential $u \brc{r}$ is a cut and
shifted Lennard-Jones potential with $u \brc{r} =  u_{\txt{LJ}} \brc{r} -
u_{\txt{LJ}} \brc{r_{c}}$ where:
\begin{equation}
  u_{\txt{LJ}} \brc{r}
=
  4 \epsilon_{\txt{ff}} \sqrbrc{\brc{\frac{\sigma_{\txt{ff}}}{r}}^{12} - \brc{\frac{\sigma_{\txt{ff}}}{r}}^{6}},
\label{eqn:LJ}
\end{equation}
and $r_{c} = \sigma_{\txt{ff}}$. Here $\epsilon_{\txt{ff}}$ is the depth of the potential well and
$\sigma_{\txt{ff}}$ is the finite distance at which the potential is zero. In
this work, the Fundamental Measure Theory (FMT) is used with the White Bear functional
\cite{hansen2006}. The weighted non-local densities are:
\begin{equation}
  \bar{\rho}_{\gamma} \brc{\vec{r}}
=
 \int d \vec{r}' \rho \brc{\vec{r}'} w^{\brc{\gamma}} \brc{|\vec{r} - \vec{r}'|}
\end{equation}
with the weight functions:
\begin{alignat}{3}
  w^{\brc{3}} \brc{r}  & = \theta \brc{r - R}           &&                                   && \\
  w^{\brc{2}} \brc{r}  & = 4 \pi R w^{\brc{1}} \brc{r}  && = 4 \pi R^{2} w^{\brc{0}} \brc{r} && =    \delta \brc{r - R} \\
  w^{\brc{V2}} \brc{r} & = 4 \pi R w^{\brc{V1}} \brc{r} && = \brc{1/r}  \delta \brc{r - R}.  &&
\end{alignat}
The excess free energy density is given by $\Phi = \Phi_{s} + \Phi_{v}$ with:
\begin{align}
  \Phi_{s} 
&=
- \bar{\rho}_{0} \ln{ \brc{1 - \bar{\rho}_{3}} }
+ \frac{\bar{\rho}_{1} \bar{\rho}_{2} }{1 - \bar{\rho}_{3}} \\
\Phi_{v}
&=
- \frac{\bar{\rho}_{V1} \cdot \bar{\rho}_{V2}}{1 - \bar{\rho}_{3}}
+ \frac{1}{36 \pi \bar{\rho}_{3}^{2} (1 - \bar{\rho}_{3})^{2}}
  \brc{\bar{\rho}_{2} - \frac{\bar{\rho}_{V2}\cdot\bar{\rho}_{V2}}{\bar{\rho}_{2}}}^{3}
  \brc{\bar{\rho}_{3}+ (1 - \bar{\rho}_{3})^{2} \ln{\brc{1 - \bar{\rho}_{3}}}}.
\end{align}
The last term that needs to be defined is the external potential, $V \brc{r}$, which is defined as:
\begin{equation}
  V \brc{\vec{r}}
=
  \rho_{s} \int d \vec{r}_{s} 
  v_{\txt{LJ}} \brc{|\vec{r} - \vec{r}_{s}|}
- v_{\txt{LJ}} \brc{r_{c}},
\end{equation}
where the integral is taken over all elements assigned to the (pore) wall. The
potential $v_{\txt{LJ}} \brc{r}$ is the same as the Lennard-Jones potential
defined in equation \ref{eqn:LJ}, but with $\epsilon_{\txt{ff}}$ replaced with
$\epsilon_{\txt{sf}}$ and  $\sigma_{\txt{ff}}$ replaced with
$\sigma_{\txt{sf}}$. 
More details about the discretization of the above equations, their numerical
implementation, and how to solve them in parallel can be found in the literature
\cite{frink2000a,frink2000b,salinger2003,douglas2003,heroux2007}. Phase
transitions are tracked using the pseudo arc length continuation algorithm of
Keller \cite{keller1977,salinger2003} which have been implemented in the LOCA
software library \cite{salinger2002}.

\subsection{Geometries}

\begin{figure}
\center
\includegraphics[]{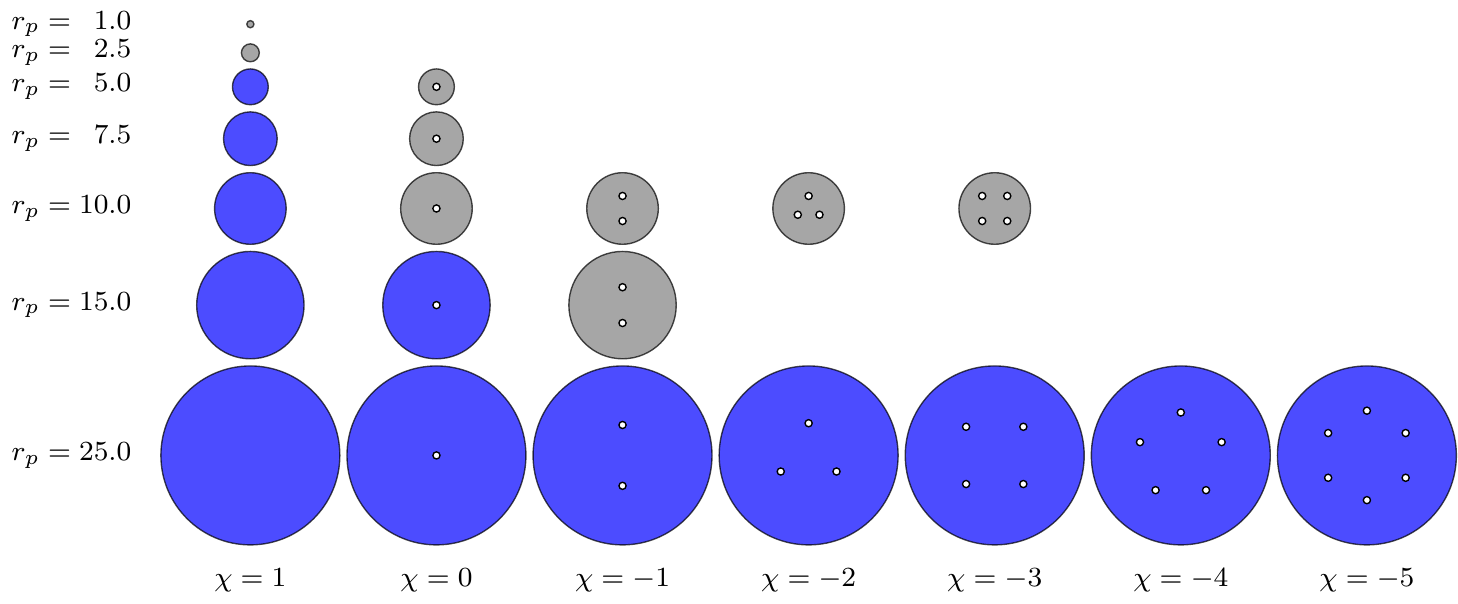}
\caption{The various geometries and topologies used in the simulations. Along
the vertical axis, the various shapes show pores with different radii, $r_{p}$.
Along the horizontal axis, rods with radius $r_{p} = 1.0$ are placed inside the
pores to modify the Euler characteristic, $\chi$. The pores in gray are
simulation cases where the distance between walls of either the pores or the
rods are smaller than $\approx 10 \sigma$, which is the distance at which Hadwiger's
theorem starts to break down \cite{konig2004}.
\label{fig:geometries}}
\end{figure}
To study how capillary condensation depends on the Minkowski functionals,
simulations have been performed for a broad range of pore sizes and topologies.
Figure~\ref{fig:geometries} shows these various geometries and topologies.
Along the vertical axis, the various shapes show pores
with different radii, $r_{p}$. Along the horizontal axis rods with radius $r_{r}
= 1.0$ are placed inside the pores to modify the Euler characteristic $\chi$.
The various 2D Minkowski functionals associated with the surface area,
circumference, and signed curvature, respectively, can be computed with the
following set of equations:
\begin{alignat}{2}
A &= \pi \; r_{p}^{2} &&- n_{r} \pi \; r_{r}^{2} \label{eqn:A}, \\
C &= \pi \; r_{p}     &&+ n_{r} \pi \; r_{r}                  , \\
K &= \pi              &&- n_{r} \pi \; \phantom{r_{r}} \label{eqn:K},
\end{alignat}
where $n_{r}$ is the number of rods inside the pore. As mentioned in section
\ref{sec:minkowski}, in order for the Minkowski functionals to accurately capture
the physics of capillary condensation, the conditions in Hadwiger's theorem
need to be met.
Considering that for very small pores the characteristic interaction length
between molecules becomes of the same order as the pore size, the additivity
constraint (Equation~\ref{eqn:additive}) is expected to break down first. In the
literature it is reported that an error of about $1\%$ is found when the system
size becomes of the order of  $\approx 10 \sigma_{\txt{ff}}$, where
$\sigma_{\txt{ff}}$ is the characteristic length scale of the interaction
potential between molecules \cite{konig2004}. For the pores in gray in
Figure~\ref{fig:geometries} the minimum distance between the walls of the pore and/or
the rods inside the pore is smaller than this distance, and Hadwiger's theorem is
expected to break down.

\begin{figure}
\center
\includegraphics[]{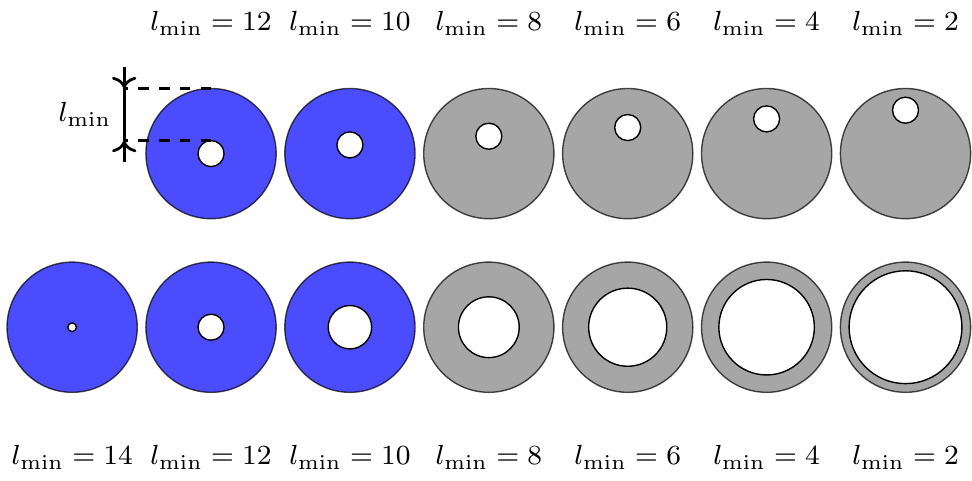}
\caption{A second set of geometries used in the simulations.
For all geometries the radius of the pore is, $r_{p} = 15.0$, and the Euler
characteristic is, $\chi = 0$. The radius of the rod in the top row geometries
is $r_{r} = 3$ and the distance with the wall is varied. In the bottom row
geometries, the radius of the rod is varied from $r_{r} = 1$, to $r_{r} = 13$. 
The pores in gray are simulation
cases in which the distance between the walls of the pore and the rod are
smaller than $\approx 10 \sigma$, the distance at which the additivity
assumption starts to break down according to literature \cite{konig2004}.
\label{fig:hadwiger}}
\end{figure}
In addition, to further explore the effect of violating Hadwiger's additivity
assumption, simulations are run for the geometries shown in figure
\ref{fig:hadwiger}. For all geometries, the radius of the pore is, $r_{p} =
15.0$, and the Euler characteristic is, $\chi = 0$. The radius of the rod in the
top row geometries is $r_{r} = 3$, and the distance with the wall is varied. In
the bottom row geometries, the radius of the rod is varied from $r_{r} = 1$, to
$r_{r} = 13$. By moving the rod towards the wall/changing the radius of the rod,
the additivity assumption gradually breaks down. This allows us to investigate
whether the accuracy of the simulations also breaks down gradually, or whether
there is a catastrophic failure at a certain wall-to-wall distance. Again, the
pores in gray are simulation cases in which the distance between the walls of
the pore and the rod are smaller than $\approx 10 \sigma_{\txt{ff}}$, the
distance at which the additivity assumption starts to break down
\cite{konig2004}.

\subsection{Simulation parameters}

\begin{table}[h]
\caption{DFT parameters of \ce{N2} and \ce{SiO2} \cite{ravikovitch1998}. The
number density for \ce{SiO2} is $\rho_{\txt{s}} = 66.15 \si{nm}^{-3}$
\cite{ustinov2005}.
Fluid-fluid interactions are truncated at $5 \sigma_{\txt{ff}}$. The simulations
are performed at $77.3 \si{K}$}
\label{tab:parameters}
\centering
\begin{tabular}{rrrrrr}
\br
                                                &
\multicolumn{1}{c}{$\epsilon_{\txt{ff}}/k_{B}$} &
\multicolumn{1}{c}{$\sigma_{\txt{ff}}$}         &
\multicolumn{1}{c}{$d_{\txt{HS}}$}              &
\multicolumn{1}{c}{$\epsilon_{\txt{sf}}/k_{B}$} &
\multicolumn{1}{c}{$\sigma_{\txt{sf}}$}         \\
                                &
\multicolumn{1}{c}{$[\si{K}]$}  &
\multicolumn{1}{c}{$[\si{nm}]$} &
\multicolumn{1}{c}{$[\si{nm}]$} &
\multicolumn{1}{c}{$[\si{K}]$}  &
\multicolumn{1}{c}{$[\si{nm}]$} \\
\mr
\ce{N2} &
94.45   &
0.3575  &
0.3575  & 
147.3   &
0.317   \\
\br
\end{tabular}
\end{table}
Existing literature has focused on the behavior of hard sphere fluids
\cite{konig2004}. This kind of potential resembles a high temperature
gas and can be a useful simplification of a system of interest. However, many
engineering applications require more complex particle-particle interactions
like the Lennard-Jones potential.
In this work, simulations have been performed for both a hard-sphere fluid and a
Lennard-Jones fluid. The hard-sphere fluid allows for a comparison with
literature and provides a simplified base case. The Lennard-Jones fluid, on the
other hand, is used to analyze the effect of adding an attractive longer range
component to the interaction potential and to see whether the framework of the
Minkowski functionals is also useful in more realistic engineering applications.

\begin{figure}
\center
\includegraphics[]{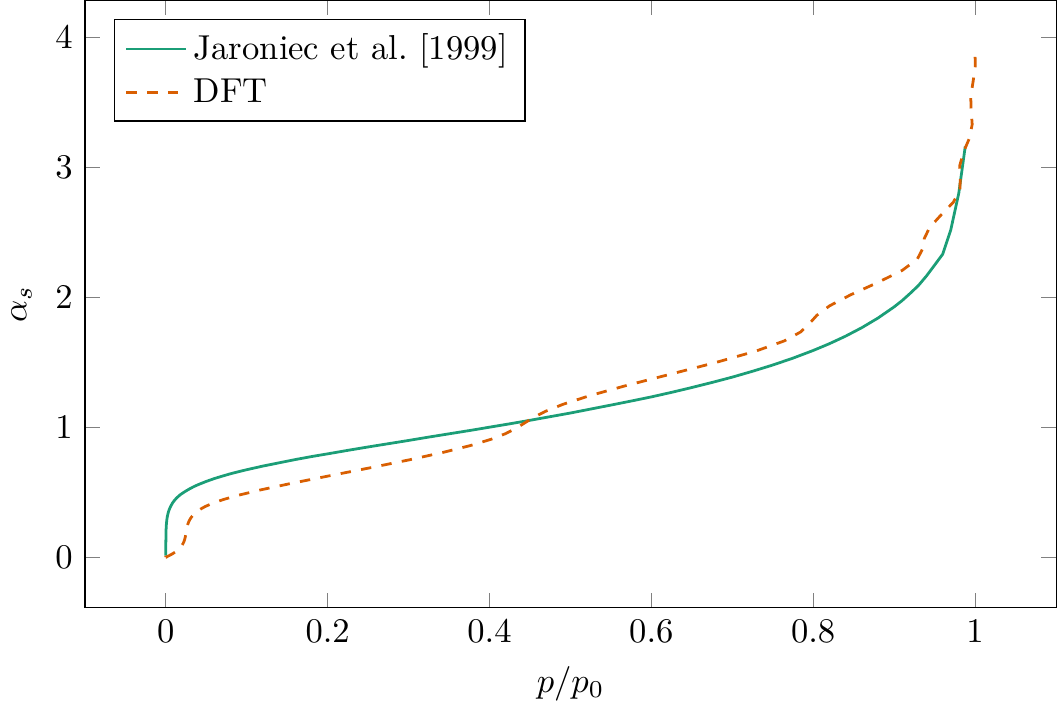}
\caption{Comparison between experiments \cite{jaroniec1999} and our DFT simulations 
for the adsorption isotherm of \ce{N2} in a \ce{SiO2} slit pore with a width of
$L = 500 \sigma_{\txt{ff}}$. The chosen wall potential does not fully
capture the interactions between \ce{N2} and \ce{SiO2}, but the results show a good
match. This confirms that the used parameters shown in table \ref{tab:parameters} are
a reasonable choice.
\label{fig:N2validation}}
\end{figure}

Table \ref{tab:parameters} shows the parameters used in the Lennard-Jones DFT simulations.
Because it is a commonly used model system
\cite{brewer1962,levitz1991,boher2014}, the Lennard-Jones fluid is parameterized
as Nitrogen in Vycor glass. The parameters are the same as those used by
Ravikovitch et al. \cite{ravikovitch1998} and Ustinov et al. \cite{ustinov2005} and are very similar to the
parameters used by Gelb \& Gubbins \cite{gelb1998} in their Grand Canonical Monte Carlo
simulations of Nitrogen in Vycor glass. 
Figure \ref{fig:N2validation} shows a comparison between DFT simulations and
experiments \cite{jaroniec1999} for the adsorption isotherm of \ce{N2} in a
\ce{SiO2} slit pore with a width of $L/\sigma_{\txt{ff}} = 500$. 
The results confirm that the parameters listed in table \ref{tab:parameters} are
a reasonable choice. With more advanced models for the interaction between
\ce{N2} and \ce{SiO2}, a
better match can be obtained between DFT simulations and experiments
\cite{ravikovitch1998,ustinov2005}. However, the choice of the same potential
for particle-particle and wall-particle keeps the system simple and the results more
easy to interpret. The computations are performed in the grand canonical
ensemble ($\mu,V, T$) and the relation between the chemical potential and
pressure was obtained from a bulk DFT simulation.

\section{Results}

In this section, the results are shown for both simulations with a hard-sphere
interaction potential and a Lennard-Jones interaction potential. All the
parameters have been made dimensionless with $\beta = k_{\txt{B}}T$ and
$\sigma_{\txt{ff}}$.

\subsection{Hard sphere potential}

\begin{figure}
\center
\includegraphics[]{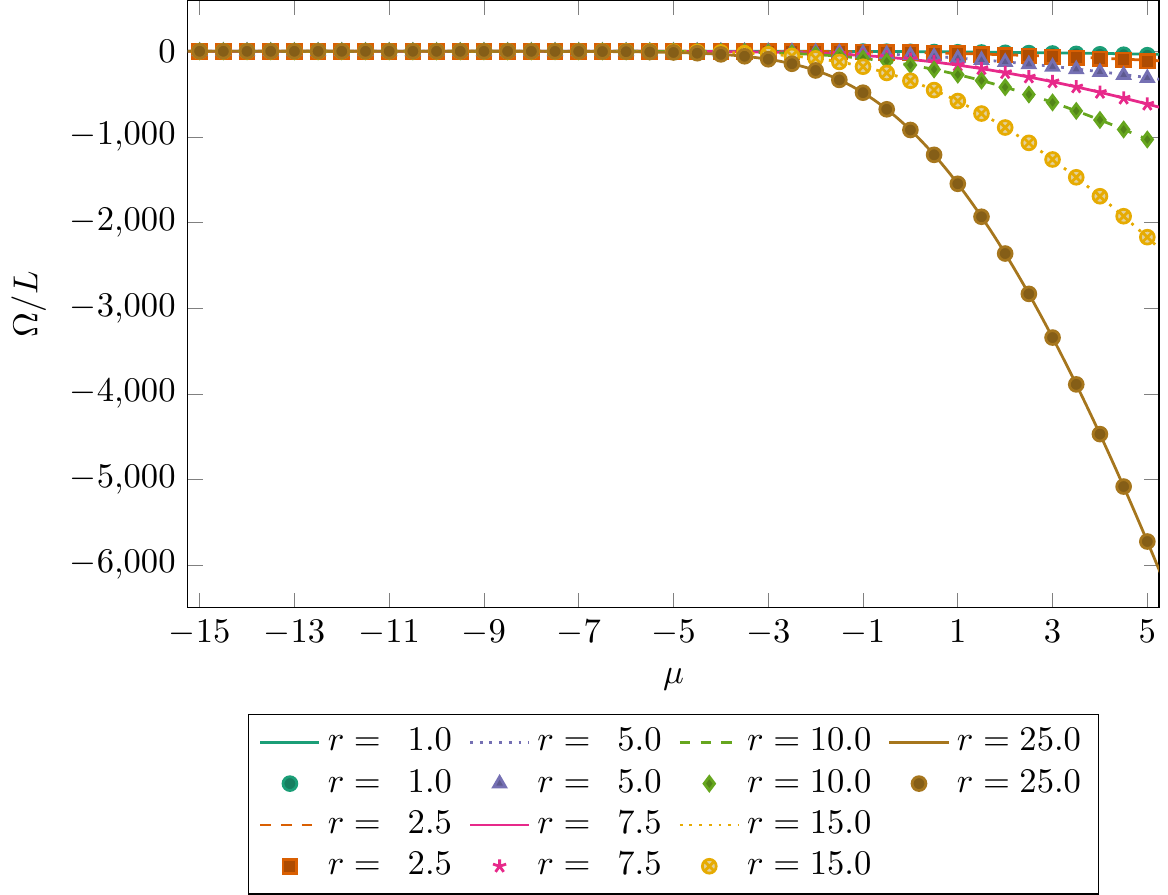}
\caption{Dimensionless 2D grand potential, $\Omega/L$, as a function of the
dimensionless chemical potential, $\mu$, for a hard sphere fluid. For clarity,
only the simulation results from Figure~\ref{fig:geometries} when $\chi = 1$ are
shown. The different lines show the results of the DFT simulations, and the
symbols show the grand potential as reconstructed from the Minkowski
functionals. The reconstruction of the grand potential uses only one set of Minkowski functional coefficients: pressure, $p \brc{\mu, T}$,
surface tension, $\sigma \brc{\mu, T}$, and bending rigidity, $\kappa \brc{\mu, T}$. 
\label{fig:omegaXp1HS}}
\end{figure}
Figure~\ref{fig:omegaXp1HS} shows the 2D dimensionless grand potential,
$\Omega/L$, as a function of the dimensionless chemical potential, $\mu$, for a
hard-sphere fluid. For clarity, only the simulation results when $\chi = 1$ are shown. The different lines show
the results of the DFT simulations while the symbols show the grand potential as
reconstructed from the Minkowski functionals and one set of Minkowski functional
coefficients: pressure, $p \brc{\mu, T}$, surface tension, $\sigma \brc{\mu, T}$,
and bending rigidity, $\kappa \brc{\mu, T}$. These coefficients are computed by
performing a least squares fit on all the simulations. The various curves show a low density regime for
chemical potentials $\mu < 0$ and a transition to a high density regime for $\mu
> 0$. Visually, in both regimes, the data sets show a good match for all
geometries.

\begin{figure}
\center
\includegraphics[]{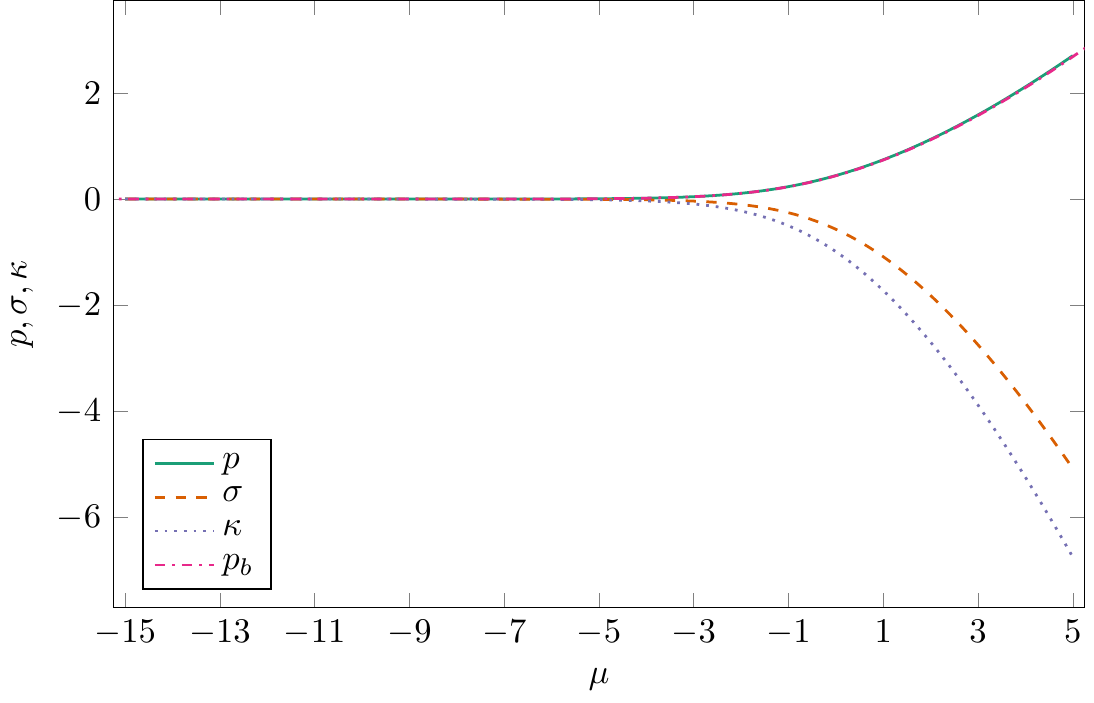}
\caption{Dimensionless Minkowski functional coefficients: pressure, $p
\brc{\mu, T}$, density, $\sigma \brc{\mu, T}$, and bending rigidity, $\kappa
\brc{\mu, T}$, as a function of the dimensionless chemical potential, $\mu$. These
are the values of the coefficients that are used in Figure~\ref{fig:omegaXp1HS}
to reconstruct the grand potential as a function of the chemical potential.
\label{fig:minkHS}}
\end{figure}
The individual dimensionless Minkowski functional coefficients are shown in
Figure~\ref{fig:minkHS} as a function of the dimensionless chemical potential,
$\mu$. These are the values of the coefficients that are used in
Figure~\ref{fig:omegaXp1HS} to reconstruct the grand potential as a function of
the chemical potential. It can be seen that in this system, the pressure
coefficient is the same as the bulk pressure. The surface tension and bending
rigidity show the same behavior as in a 3D system \cite{konig2004}.

\begin{figure}
\center
\includegraphics[]{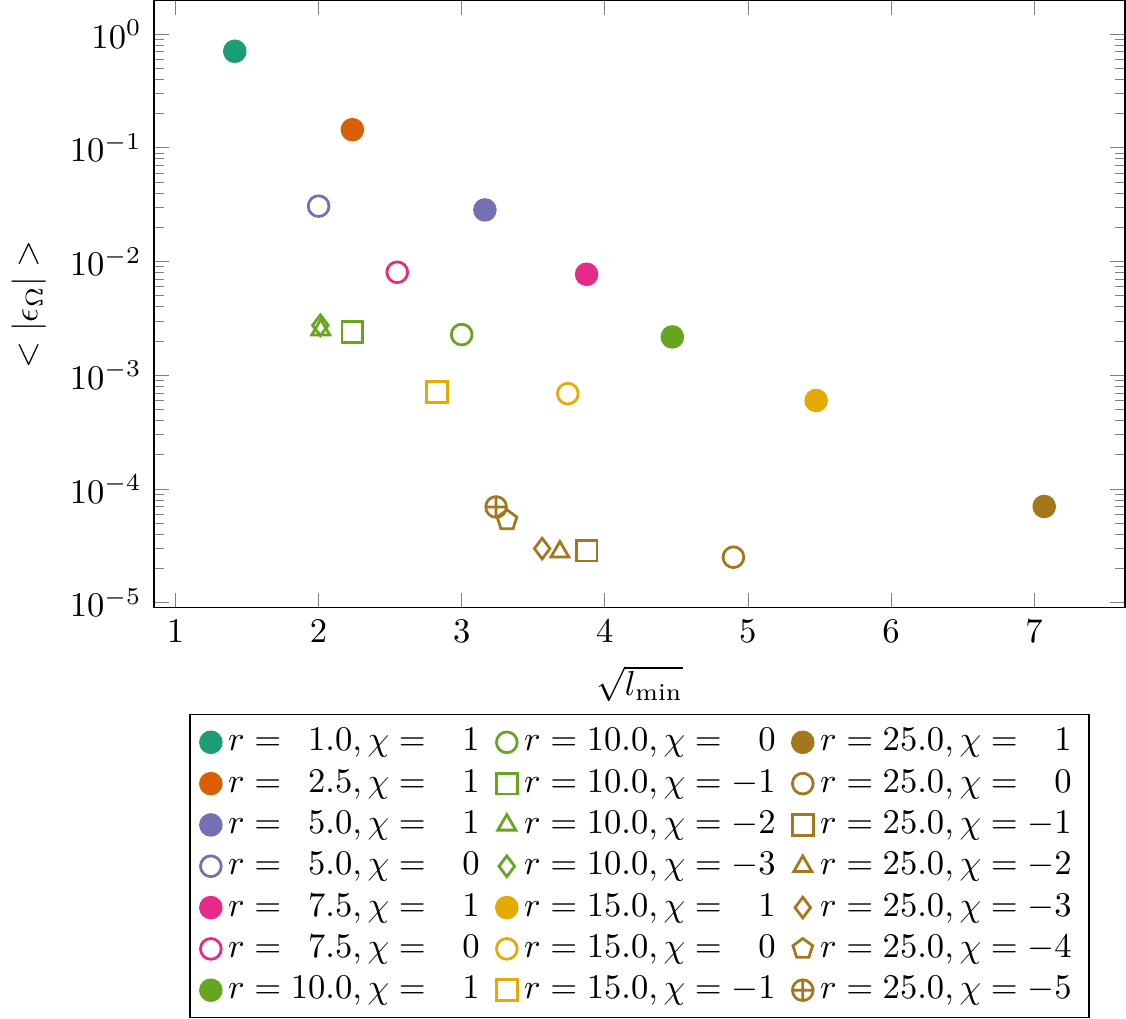}
\caption{Average absolute relative error, $<\abs{\epsilon_{\Omega}}>$, as a
function of the square root of the minimal characteristic length scale of the
system, $\sqrt{l_{\txt{min}}}$. In the case of a pore without rods,
this distance is twice the radius. When rods are present within the pore, this is
the smallest distance between the pore wall and a rod or between two different
rods. 
\label{fig:omegaErrHS}}
\end{figure}
To get a better understanding of the error introduced by using the Minkowski
functionals, Figure~\ref{fig:omegaErrHS} shows the average absolute relative
error, $<\abs{\epsilon_{\Omega}}>$, as a function of the minimal characteristic length scale
of the system, $l_{\txt{min}}$. The average absolute relative error is defined as:
\begin{equation}
  \avr{ \abs{\epsilon_{\Omega}} }
=
  \avr{ \abs{\frac{\Omega_{\txt{DFT}} - \Omega_{\txt{Mink}}}{ \Omega_{\txt{DFT}} }} }_{\mu},
\label{eqn:error}
\end{equation}
where $\Omega_{\txt{DFT}}$ is the grand potential computed using DFT, and
$\Omega_{\txt{Mink}}$ is the reconstruction of the grand potential using the
Minkowski functionals. The error is averaged with respect to the chemical
potential, $\mu$. In the case of a pore without rods,
the length scale, $l_{\txt{min}}$, equals twice the radius. When rods are
present within the pore, this is
the smallest distance between the pore wall and a rod or between two different
rods. For pores without rods, it can be observed that
$\log{(<\abs{\epsilon_{\Omega}}>)}$ scales approximately linearly with
$\sqrt{l_{\txt{min}}}$. 
Figure~\ref{fig:omegaXp1HS} indicates that there is a very good
match between the grand potential computed directly using DFT and the
reconstruction using the Minkowski functionals;  nevertheless, close inspection
of the results shows that the accuracy of the Minkowski functionals declines as
the pores become smaller. This is due to the fact that when the system size is
of the same order as the interaction length between molecules, Hadwiger's
assumption of additivity breaks down. However, looking at other topologies
suggests that $l_{\txt{min}}$ is not a perfect description of the characteristic
length scale of the system. Changing the topology of the system with rods seems
to have little effect on the error and the linear scaling relation between
$\sqrt{l_{\txt{min}}}$ and $\log{(<\abs{\epsilon_{\Omega}}>)}$ does not
hold. The error, $\avr{ \abs{\epsilon_{\Omega}} }$, found in this work for cylindrical pores is consistent with the
literature \cite{konig2004}.

\begin{figure}
\center
\includegraphics[]{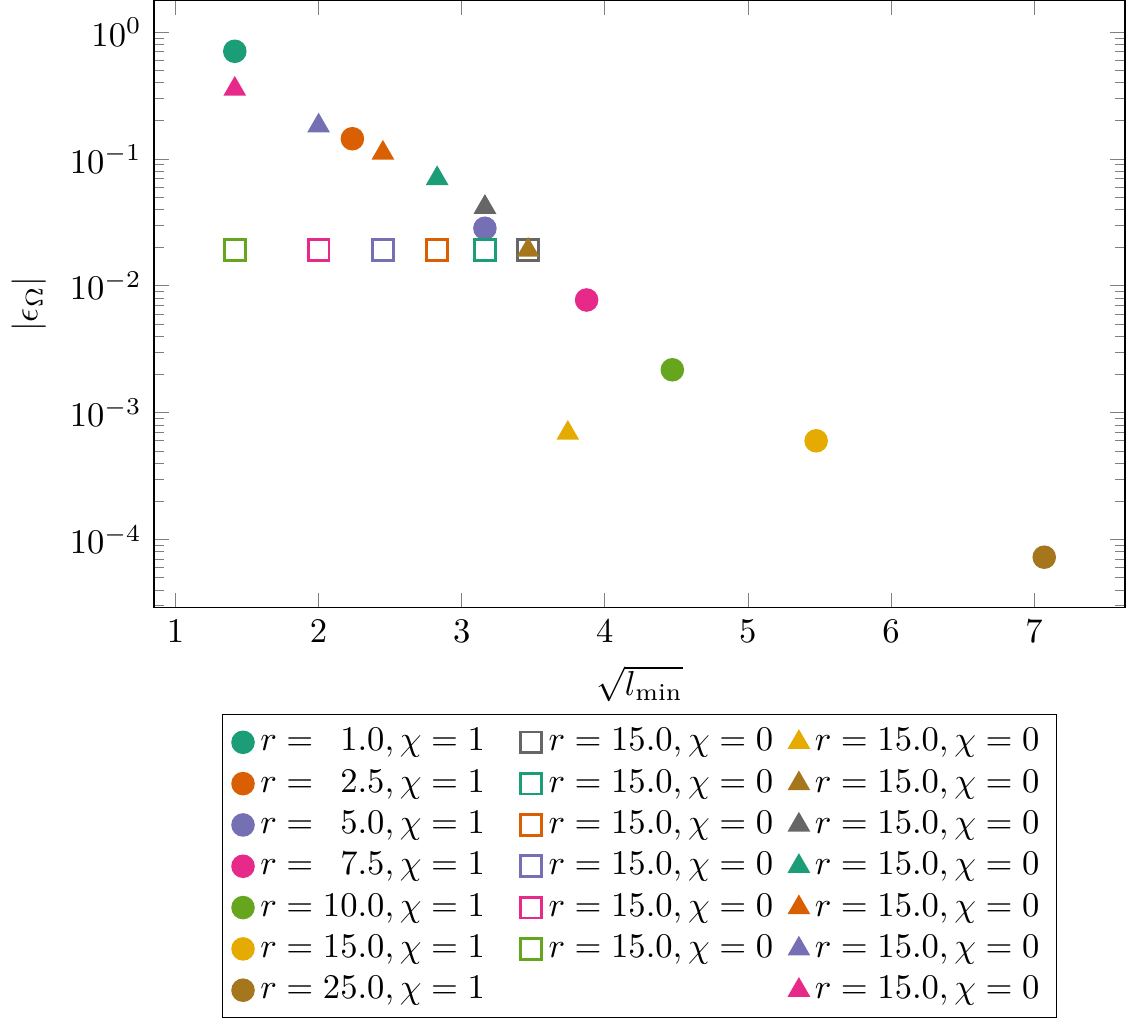}
\caption{Relative error $\abs{\epsilon_{\Omega}}$ as a function of the minimal
characteristic length scale of the system, $l_{\txt{min}}$. In the case of a
pore without rods,
this distance is twice the radius. When rods are present within the pore this is
the smallest distance between the pore wall and a rod or between two different
rods. 
\label{fig:omegaErrSR}}
\end{figure}
To further analyze the behavior of the error, $\avr{ \abs{\epsilon_{\Omega}} }$,
as a function of various morphologies
and topologies, Figure~\ref{fig:omegaErrSR} shows an analysis of the error when
applying the Minkowski functional expansion of the grand potential for the
geometries shown in Figure~\ref{fig:hadwiger}. The top row of this figure shows
a set of pore geometries with one rod inside. In these different geometries, the
distance from the pore to the wall is varied. All of
these geometries have the same Minkowski functionals. The bottom row of
Figure~\ref{fig:hadwiger} also shows a set of pore geometries with one rod
inside, but the pore is centered in the middle. In this set of geometries, the
size of the rod is varied.
Figure~\ref{fig:omegaErrSR} shows the average absolute relative error,
$<\abs{\epsilon_{\Omega}}>$, as a function of the minimal characteristic length
scale of the system, $l_{\txt{min}}$. The closed symbols show reference DFT
simulations of pores without rods, open squares show simulations from the top
row of Figure~\ref{fig:hadwiger}, and closed triangles show results from the
bottom row of Figure~\ref{fig:hadwiger}. The figure shows that the scaling of
the error as a function of the minimal distance, $l_{\txt{min}}$, is very similar
for open pores and pores with rods of varying radii. However, moving a rod
around inside a pore does not have much effect on the error of the Minkowski
functional reconstruction of the grand potential. These results confirm that
$l_{\txt{min}}$ is not a perfect description of the characteristic length scale
of the system. In addition, the literature suggests that perturbations introduced at
a caustic point should increase the error \cite{konig2004}, which is not the
case for these simulations. The error was also studied as a function of the length scale
defined as the ratio of the Minkowski functionals for surface area and
circumference $A \brc{X}/C \brc{X}$ and as the average wall-to-wall distance.
However, neither of these measures improved the scaling relation and more study
is needed.

\begin{figure}
\center
\includegraphics[]{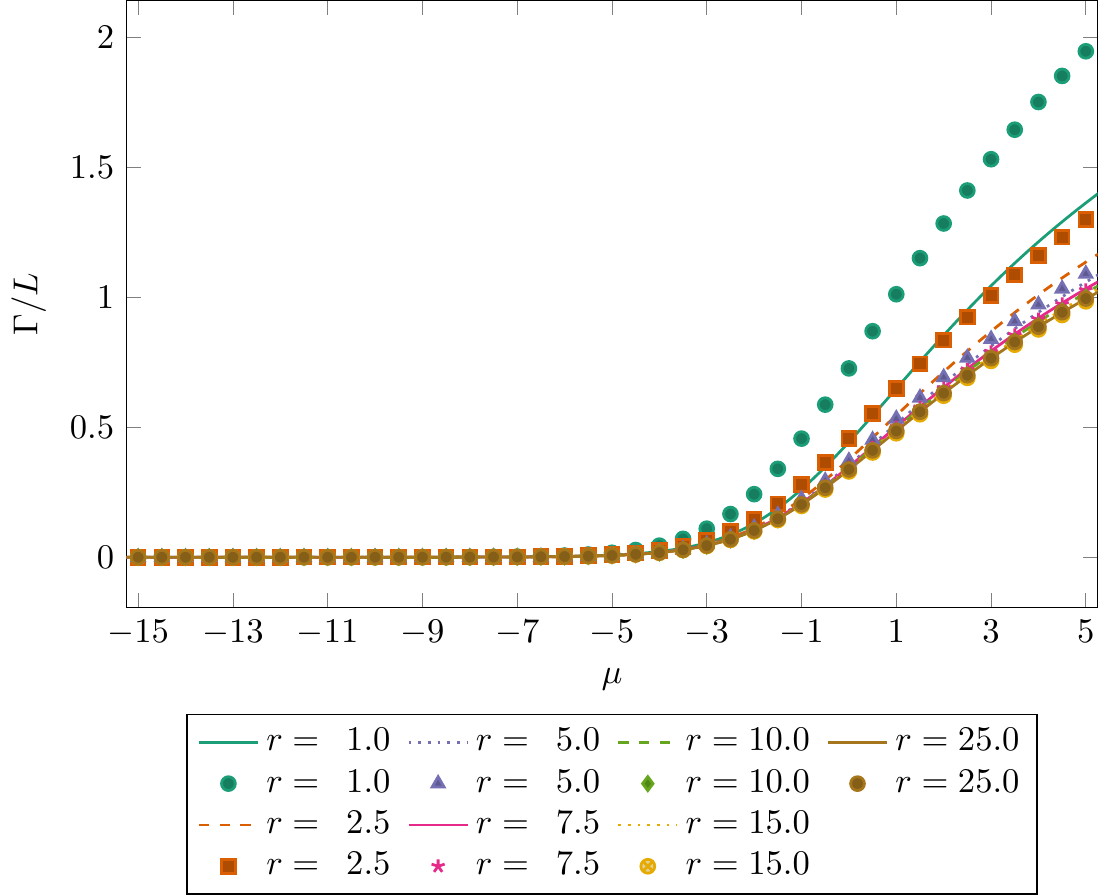}
\caption{Dimensionless 2D excess adsorption, $\Gamma/L$, as a function of the
dimensionless chemical potential, $\mu$, for a hard sphere fluid. For clarity,
only the simulation results from Figure~\ref{fig:geometries} when $\chi = 1$ are
shown. The different lines show the results of the DFT simulations while the
symbols show the grand potential as reconstructed from the Minkowski functionals
and one set of Minkowski functional coefficients: the derivatives of
pressure,$\partial (p-p_{b})/{\partial \mu}$, surface tension, ${\partial
\sigma}/{\partial \mu}$, and bending rigidity, ${\partial \kappa}/{\partial
\mu}$, with respect to the chemical potential. 
\label{fig:gammaXp1HS}}
\end{figure}

An important thermodynamic property which can be derived from the grand
potential is the excess adsorption, $\Gamma$. Figure \ref{fig:gammaXp1HS} shows
the 2D dimensionless excess adsorption, $\Gamma/L$, as a function of the chemical
potential, $\mu$. Again, only the simulation results from
Figure~\ref{fig:geometries} when $\chi = 1$ are shown. The different lines show
the results of the DFT simulations. The symbols show the excess adsorption
as reconstructed from the Minkowski functionals and one set of Minkowski
functional coefficients: the derivatives of pressure,$\partial
(p-p_{b})/{\partial \mu}$, surface tension, ${\partial \sigma}/{\partial \mu}$,
 and bending rigidity, ${\partial \kappa}/{\partial \mu}$, with respect to the
chemical potential. Like the grand potential, these coefficients are computed by
performing a least squares fit on all the simulations from
Figure~\ref{fig:geometries}. For larger pores there is a very good match between the results of
the DFT computations and the reconstruction of the adsorption isotherm using
Minkowski functionals. However, for smaller pores a clear difference can be
observed. This difference between the DFT simulations and the Minkowski
functional reconstruction is more pronounced than for the grand
potential in Figure~\ref{fig:omegaXp1HS}. This can most likely be contributed to
the fact that the excess adsorption is a derivative of the grand potential,
which introduces additional uncertainty in the results. A second observation is
the collapse of the different adsorption isotherms for larger pore sizes.
Since the difference between the pressure, $p$, and the bulk pressure, $p_{b}$,
is close to zero (see Figure~\ref{fig:minkHS}), the only term in
Equation~\ref{eqn:excess} that can contribute to differences in excess adsorption between
different geometries is $K \brc{X}/C \brc{X}$. This suggests that, as the pore
size increases, the effect of topology on the excess adsorption decreases.

\begin{figure}
\center
\includegraphics[]{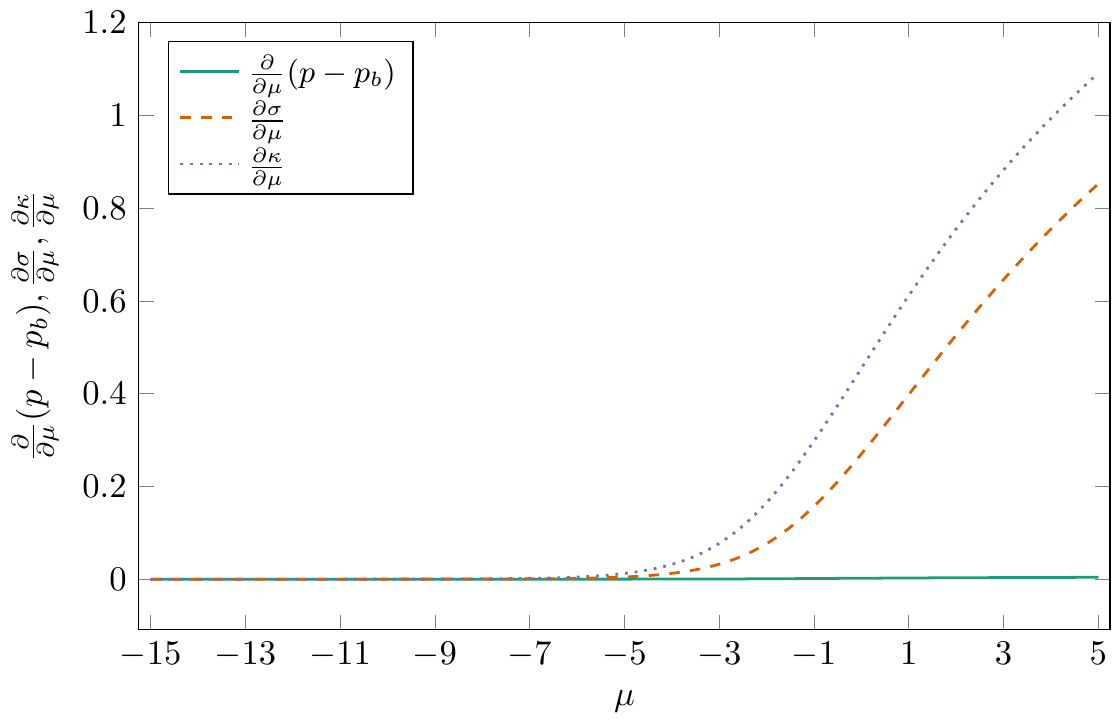}
\caption{Dimensionless Minkowski functional coefficients: the derivatives of
pressure,$\partial (p-p_{b})/{\partial \mu}$, surface tension, ${\partial
\sigma}/{\partial \mu}$, and bending rigidity, ${\partial \kappa}/{\partial
\mu}$, with respect to the chemical potential. These are the values of the
coefficients that are used in Figure~\ref{fig:gammaXp1HS} to reconstruct the
excess adsorption as a function of the chemical potential. 
\label{fig:dMinkHS}}
\end{figure}
The matching dimensionless Minkowski functional coefficients: the derivatives of
pressure,$\partial (p-p_{b})/{\partial \mu}$, surface tension, ${\partial
\sigma}/{\partial \mu}$, and bending rigidity, ${\partial \kappa}/{\partial
\mu}$, with respect to the chemical potential are shown in
Figure~\ref{fig:dMinkHS}. This figure confirms again that in this system the
pressure coefficient is the same as the bulk pressure. Considering the the minus
sign in Equation~\ref{eqn:excess}, the curves for the surface tension and bending
rigidity are consistent with Figure~\ref{fig:minkHS}.

\begin{figure}
\center
\includegraphics[]{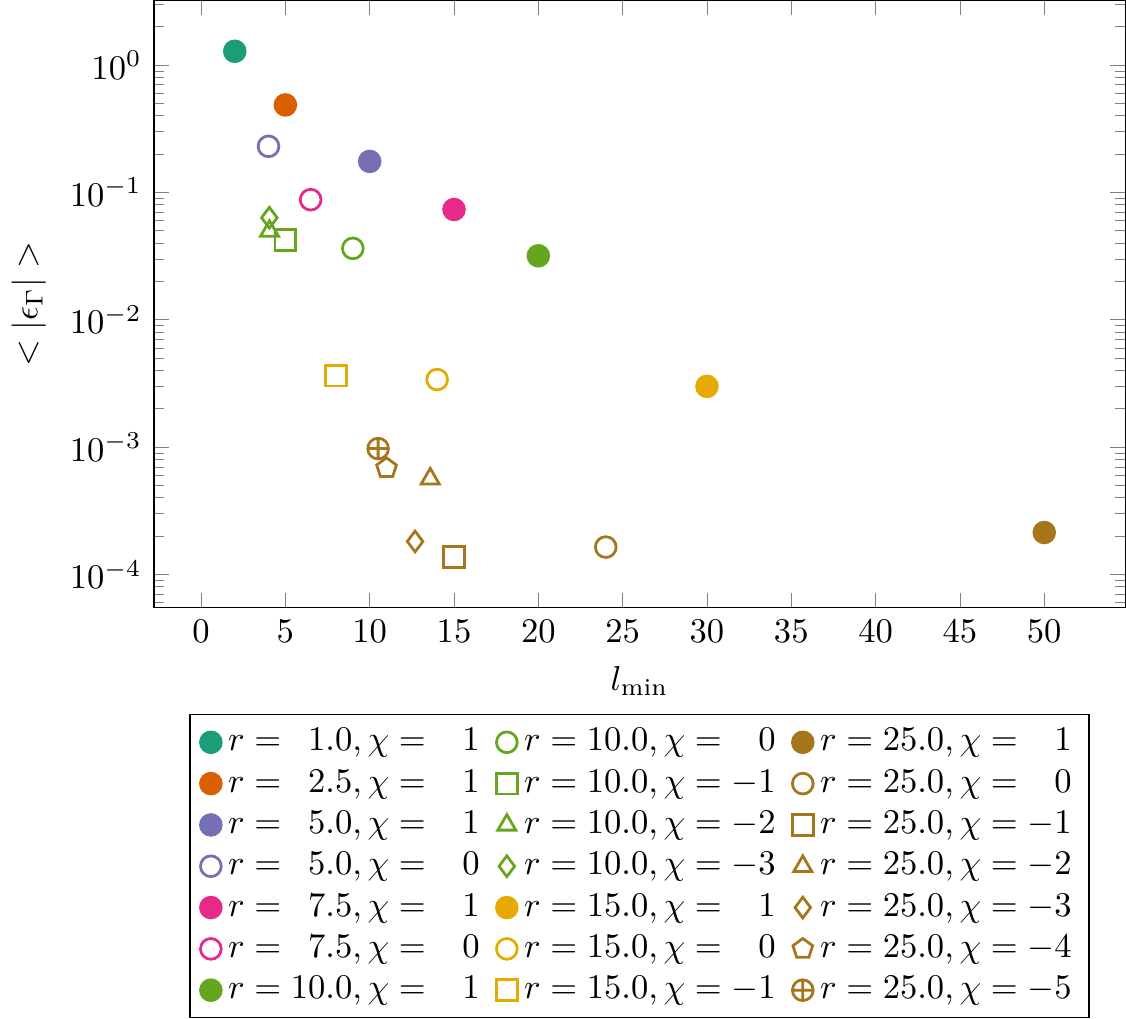}
\caption{Average absolute relative error, $<\abs{\epsilon_{\Gamma}}>$, as a function of the minimal
characteristic length scale of the system, $l_{\txt{min}}$. In the case of a
pore without rods, this distance is twice the radius. When rods are present within the pore this is
the smallest distance between the pore wall and a rod or between two different
rods. 
\label{fig:gammaErrHS}}
\end{figure}

The last graph for the hard-sphere system is Figure~\ref{fig:gammaErrHS}, which
shows the average absolute relative error $<\abs{\epsilon_{\Gamma}}>$ as a function of the minimal
characteristic length scale of the system, $l_{\txt{min}}$. In the case of a
pore without rods,
this distance is twice the radius. When rods are present within the pore, this is
the smallest distance between the pore wall and a rod or between two different
rods. The relative error is defined in the same manner as in
equation~\ref{eqn:error}. For pores without rods, it can be observed that
$\log{(<\abs{\epsilon_{\Gamma}}>)}$ scales almost linearly with $l_{\txt{min}}$
instead of being proportional to $\sqrt{l_{\txt{min}}}$. In addition, due to the
fact that the excess adsorption is a derivative of the grand potential, the
error is larger than observed in Figure~\ref{fig:omegaErrHS}. 

\subsection{Lennard-Jones potential}

In this section, the results for the Lennard-Jones fluid are presented. Due to
the increased interaction length and the more complex phase behavior of this
interaction potential compared to a hard-sphere fluid, it is found that Hadwiger's theorem
starts to break down earlier. However, as is shown below, by performing a scaling
analysis of the phase transitions of the system as a function of the Minkowski
functionals, the following additional terms $p' \brc{\mu, T} A^{1/2} \brc{X}$
and $\sigma' \brc{\mu, T} C^{3/4} \brc{X}$ can be identified and added to the
expansion of the grand potential for increased accuracy. All the results in this
section include these extra terms. 

\begin{figure}
\center
\includegraphics[]{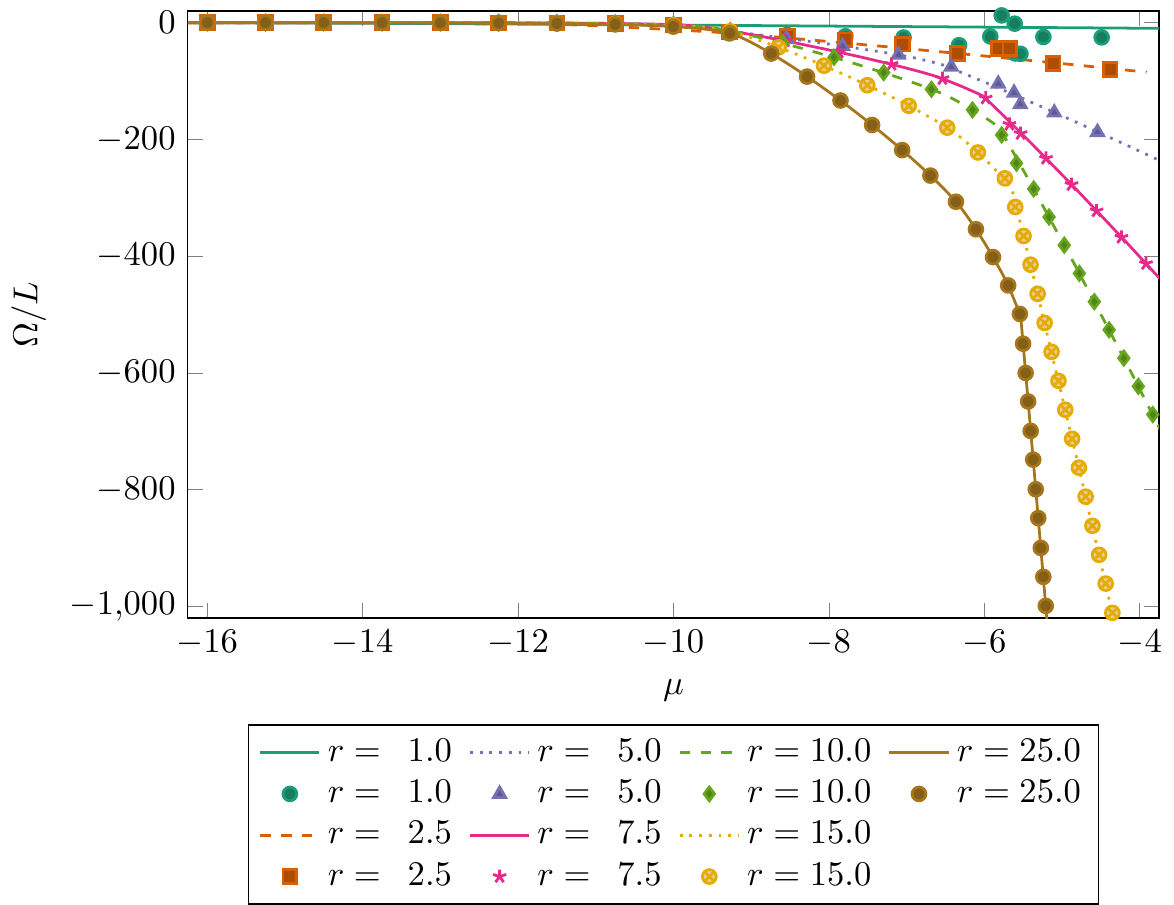}
\caption{Dimensionless 2D grand potential, $\Omega/L$, as a function of the
dimensionless chemical potential, $\mu$, for a Lennard-Jones fluid. Only the
simulation results when $\chi = 1$ are
shown. The different lines show the results of the DFT simulations while the
symbols show the grand potential as reconstructed from the Minkowski functionals
and one set of Minkowski functional coefficients: pressure, $p \brc{\mu, T}$,
surface tension, $\sigma \brc{\mu, T}$, bending rigidity, $\kappa \brc{\mu, T}$,
and the pseudo pressure and surface tension terms $p' \brc{\mu, T}$ and $\sigma'
\brc{\mu, T}$. 
\label{fig:omegaXp1LJ}}
\end{figure}
Figure~\ref{fig:omegaXp1LJ} shows the dimensionless 2D grand potential,
$\Omega/L$, as a function of the dimensionless chemical potential, $\mu$, for a
Lennard-Jones fluid. Only the simulation results with Euler characteristic,
$\chi = 1$, are shown. The different lines show
the results of the DFT simulations while the symbols show the grand potential 
reconstructed from the Minkowski functionals and one set of Minkowski functional
coefficients: pressure, $p \brc{\mu, T}$, surface tension, $\sigma \brc{\mu, T}$, bending rigidity, $\kappa
\brc{\mu, T}$, and the pseudo pressure and surface tension terms $p' \brc{\mu,
T}$ and $\sigma' \brc{\mu, T}$. The shape of the grand potential curves shows
more complex behavior than those in Figure~\ref{fig:omegaXp1HS} for a
hard-sphere fluid. For the pore size, $r_{p} = 25$, three different regimes can
be identified: i) at low chemical potential the pores are completely empty, ii) starting
at about $\mu \approx -10$, a second-order phase transition can be observed and
gas starts adsorbing on the wall, and iii) at about $\mu \approx -6$ capillary
condensation can be observed. In this work, capillary condensation is defined as
a first-order phase transition that can be identified by a van der Waals loop
in the grand potential as a function of the chemical potential \cite{evans1987}.
The data sets for DFT simulations and the reconstructions using the Minkowski
functionals show a good match for larger geometries, but a discrepancy for the
smallest pores. Like the hard-sphere fluid, this is the consequence of the
breakdown of Hadwiger's theorem. One example of how this manifests itself, is
that there is a critical pore size at which the first order capillary condensation phase
transition changes into a second order transition. It is found that this change
from a first-order to a second-order phase transition is not captured well by
the Minkowski functional approximation of the grand potential.

\begin{figure}
\center
\includegraphics[]{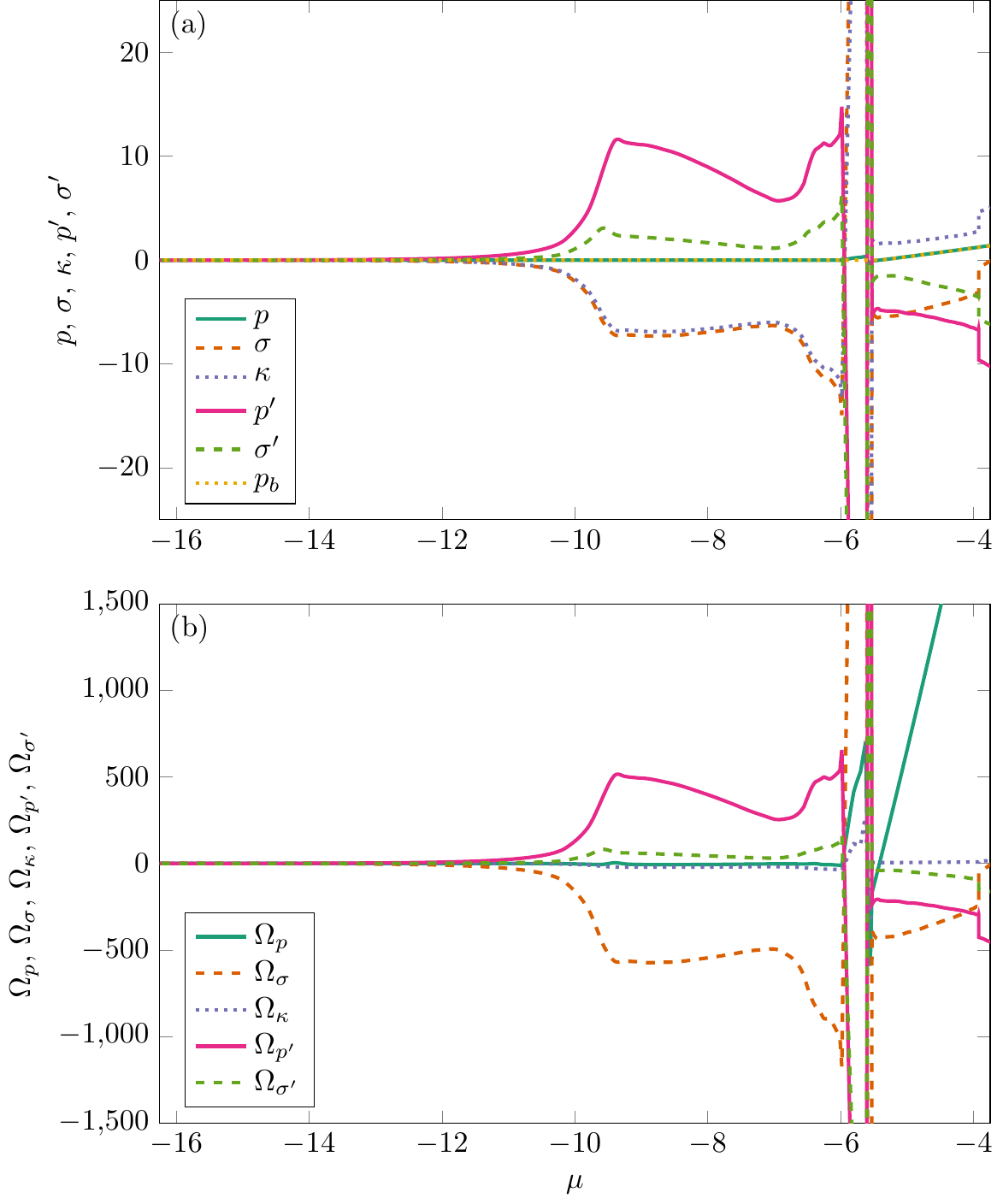}
\caption{(a) Dimensionless Minkowski functional coefficients: pressure, $p \brc{\mu,
T}$, density, $\sigma \brc{\mu, T}$, bending rigidity, $\kappa \brc{\mu, T}$,
and the pseudo pressure and surface tension terms $p' \brc{\mu, T}$ and $\sigma'
\brc{\mu, T}$, as a function of the dimensionless chemical potential, $\mu$. These
are the values of the coefficients that are used in Figure~\ref{fig:omegaXp1LJ}
to reconstruct the grand potential as a function of the chemical potential. Like the hard-sphere system, also in this system the pressure
coefficient is very similar to the bulk pressure, $p_{b}$. 
\label{fig:minkLJ}}
\end{figure}
The corresponding dimensionless Minkowski functional coefficients as a function of
the dimensionless chemical potential, $\mu$, are shown in
Figure~\ref{fig:minkLJ}~(a). It can be observed that like the hard-sphere
system, the pressure coefficient is very similar to the bulk pressure, $p_{b}$.
All coefficients show a large peak around $\mu \approx -6$ and the range of these peaks
extends from about $-150$ to $150$. Before these peaks occur, the surface tension
and bending rigidity terms show very similar behavior.
To get a better understanding of the behavior of a Lennard-Jones fluid under
confinement, one can analyze how much individual Minkowski functionals contribute
to the grand potential.
In Figure~\ref{fig:minkLJ}~(b), the value of the Minkowski functional coefficients
times their corresponding Minkowski functionals is shown for a pore with radius,
$r_{p} = 25$, and without any rods inside, $\chi = 1$. This analysis suggests
that the adsorption of gas onto the pore wall is dominated by the surface
tension and the pseudo pressure. When capillary condensation occurs, the surface
tension and pseudo pressure contributions both become discontinuous and show
large increases. Also, the contribution from the pseudo surface tension
contribution becomes significant. After the bulk phase transition, the system becomes
increasingly dominated by the pressure. As expected, there is no significant
contribution from the topology in this pore geometry.

\begin{figure}
\center
\includegraphics[]{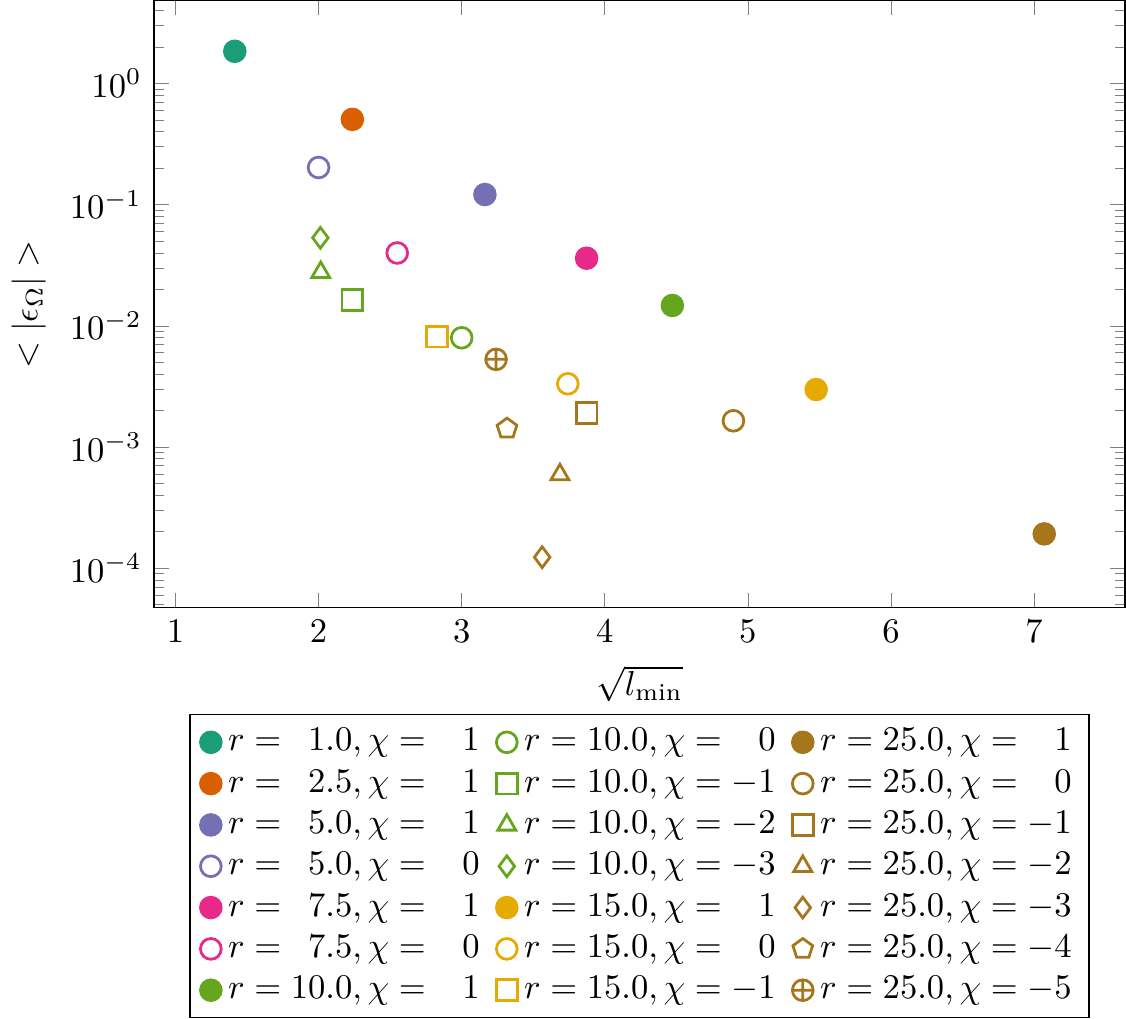}
\caption{Average absolute relative error $<\abs{\epsilon_{\Omega}}>$ as a function
of the minimal characteristic length scale of the system, $l_{\txt{min}}$. In
the case of a pore without rods, this distance is twice the radius. When rods are
present within the pore this is the smallest distance between the pore wall and
a rod or between two different rods. 
\label{fig:omegaErrLJ}}
\end{figure}
Figure~\ref{fig:omegaErrLJ} shows the average absolute relative error,
$<\abs{\epsilon_{\Omega}}>$, as a function of the square root of the minimal
characteristic length scale of the system, $\sqrt{l_{\txt{min}}}$. As a
reminder, the average absolute relative error is defined as:
\begin{equation}
  \avr{ \abs{\epsilon_{\Omega}} }
=
  \avr{ \abs{\frac{\Omega_{\txt{DFT}} - \Omega_{\txt{Mink}}}{ \Omega_{\txt{DFT}} }} }_{\mu},
\label{eqn:error}
\end{equation}
where $\Omega_{\txt{DFT}}$ is the grand potential computed using DFT, and
$\Omega_{\txt{Mink}}$ is the reconstruction of the grand potential using the
Minkowski functionals. The error is averaged with respect to the chemical
potential, $\mu$. Due to the longer interaction length of
the Lennard-Jones potential compared to the hard-sphere potential and the more
complex phase behavior, the observed error is larger than in
Figure~\ref{fig:omegaErrHS}. While the cut-off length of the interaction
potential is equal to $5 \sigma$ in the DFT simulations, one could expect the
error to be significantly less than 5 times as large because the attractive tail
of the Lennard-Jones potential is almost zero at $2 \sigma$. The error is indeed
significantly less than 5 times as large. However, this is in part due to the
addition of the pseudo pressure and surface tension terms to the Minkowski
functional expression for the grand potential.
Like the hard-sphere fluid, for
pores without rods, it can be observed that $\log{(<\abs{\epsilon_{\Omega}}>)}$
scales almost linearly with $\sqrt{l_{\txt{min}}}$. However, this scaling does not hold
for pores with rods. Like the hard-sphere system, changing the topology of the
system with rods does not have a large effect on the error, $<\abs{\epsilon_{\Omega}}>$, especially for smaller
pores. The fact that, as shown above, the topology does not have a very large contribution to the
grand potential could partly explain this observation. The figure confirms the
observation in Figure~\ref{fig:omegaXp1LJ} that for larger pores the fit of the
Minkowski functional reconstruction of the grand potential is much better than
for smaller pores.

\begin{figure}
\center
\includegraphics[]{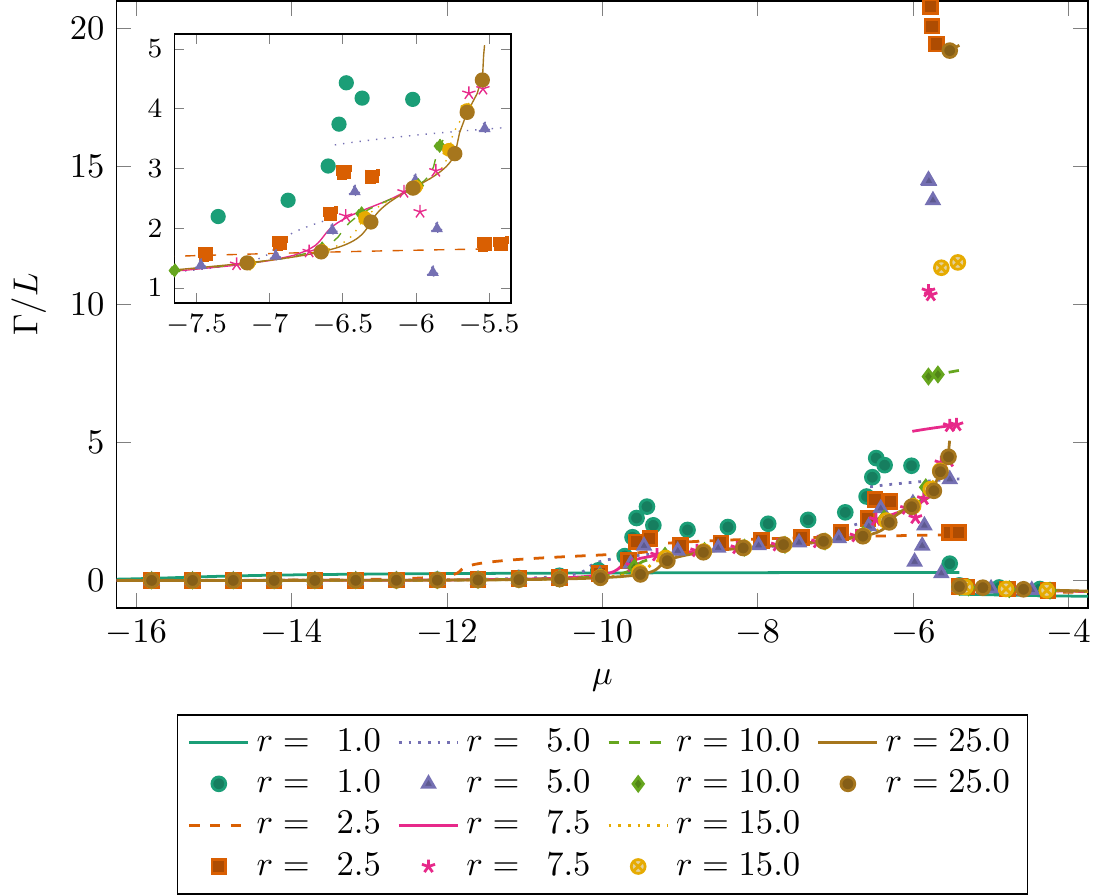}
\caption{Dimensionless 2D excess adsorption, $\Gamma/L$, as a function of the
dimensionless chemical potential, $\mu$, for a Lennard-Jones fluid. Only the
simulation results from Figure~\ref{fig:geometries} when $\chi = 1$ are
shown. The different lines show the results of the DFT simulations while the
symbols show the grand potential as reconstructed from the Minkowski functionals
and one set of Minkowski functional coefficients: the derivatives of
pressure,$\partial (p-p_{b})/{\partial \mu}$, pressure per surface area, $p'
\brc{\mu, T}$, surface tension, ${\partial \sigma}/{\partial \mu}$, bending
rigidity, ${\partial \kappa}/{\partial \mu}$, and the pseudo pressure and
surface tension terms $\partial p'/\partial \mu$ and $\partial \sigma'/\partial
\mu$ with respect to the chemical
potential. 
\label{fig:gammaXp1LJ}}
\end{figure}
The dimensionless 2D excess adsorption, $\Gamma/L$, as a function of the
dimensionless chemical potential, $\mu$, for a Lennard-Jones fluid is shown in
Figure~\ref{fig:gammaXp1LJ}. Again, only the simulation results when $\chi = 1$ are
shown. The different lines show the results of the DFT simulations while the
symbols show the grand potential as reconstructed from the Minkowski functionals
and one set of Minkowski functional coefficients: the derivatives of
pressure,$\partial (p-p_{b})/{\partial \mu}$, surface tension, ${\partial \sigma}/{\partial \mu}$, bending
rigidity, ${\partial \kappa}/{\partial \mu}$, and the pseudo pressure and
surface tension terms $\partial p'/\partial \mu$ and $\partial \sigma'/\partial
\mu$ with respect to the chemical potential. As is the case for the grand
potential in Figure~\ref{fig:omegaXp1LJ}, different regimes can be identified in
Figure~\ref{fig:gammaXp1LJ}. For the pore size, $r_{p} = 25$, the following
regimes can be observed: i) at low chemical potential the pores are completely
empty, ii) starting at about $\mu \approx -10$, a second-order phase transition
can be observed and gas starts adsorbing on the wall, iii) at about $\mu
\approx -6$ capillary condensation can be observed, and iv) around $\mu \approx
-5$ the bulk phase transition occurs and the excess adsorption shows a
significant drop. The inset shows that for larger pores there is a very good match
between the DFT computations and the Minkowski functional reconstructions.
However, for the smallest pores the match is quite poor and neither the film
adsorption stage or capillary condensation is captured well.

\begin{figure}
\center
\includegraphics[]{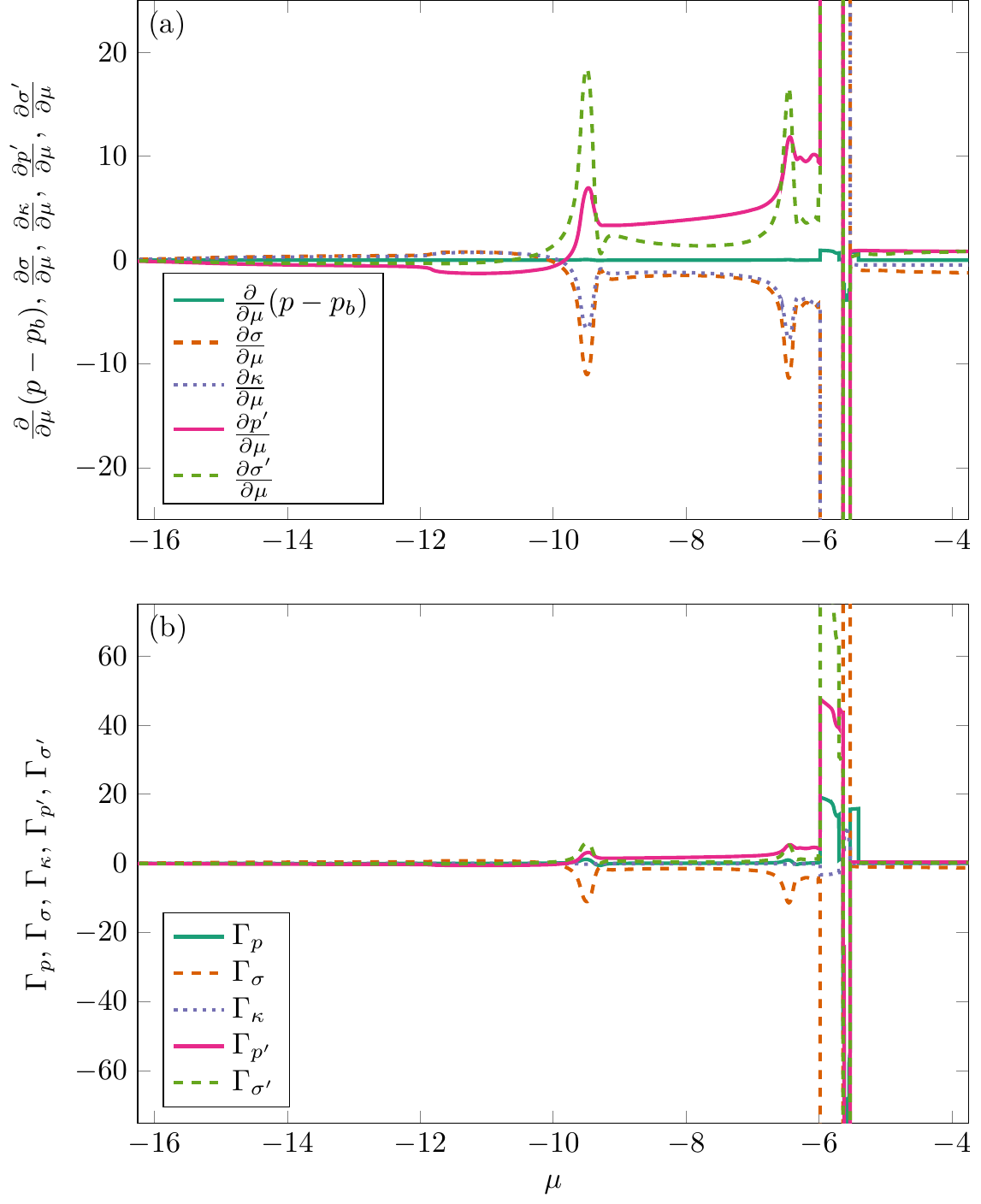}
\caption{(a) Dimensionless Minkowski functional coefficients: the derivatives of
pressure,$\partial (p-p_{b})/{\partial \mu}$, pressure per surface area, $p'
\brc{\mu, T}$,surface tension, ${\partial \sigma}/{\partial \mu}$, bending
rigidity, ${\partial \kappa}/{\partial \mu}$, and the pseudo pressure and
surface tension terms $\partial p'/\partial \mu$ and $\partial \sigma'/\partial
\mu$, with respect to the chemical
potential. These are the values of the coefficients that are used in
Figure~\ref{fig:gammaXp1LJ} to reconstruct the excess adsorption as a function of
the chemical potential. 
(b) Contribution of the Minkowski functional coefficients to the excess
adsorption for a pore with a radius of $r_{p} = 25$ and an Euler characteristic
of $\chi = 1$. 
\label{fig:dMinkLJ}}
\end{figure}

The derivatives of the Minkowski functional coefficient with respect to the
chemical potential, $\mu$, used to reconstruct the excess adsorption, can be
observed in Figure~\ref{fig:dMinkLJ}~{a}. This plot confirms that the pressure
term is very similar to the bulk pressure. The peaks at about $\mu \approx -10$
and $\mu \approx -7$ are the locations of the second-order phase transitions
associated with layers of gas molecules adsorbing onto the pore wall. To be able
to analyze how the different terms contribute to the excess adsorption, in
Figure~\ref{fig:dMinkLJ}~(b) the value of the Minkowski functional coefficients
times their corresponding Minkowski functionals is shown for a pore with radius
$r_{p} = 25$ and without any rods inside, $\chi = 1$. Since these coefficients
are derivatives of the coefficients used to reconstruct the grand potential,
they can be both positive and negative. The amount of gas adsorbed onto the
wall is dominated by the pseudo pressure and a negative contribution from the
surface tension. The pseudo surface tension only contributes during the
second-order phase transitions at about $\mu \approx -10$ and $\mu \approx -7$.
Capillary condensation is characterized by a much larger contribution of the
pseudo surface tension and many discontinuities in all the different terms to
accommodate the discontinuity of a first-order phase transition. As expected for
an open-pore geometry, both the pressure and topology do not significantly contribute to
the excess adsorption.

\begin{figure}
\center
\includegraphics[]{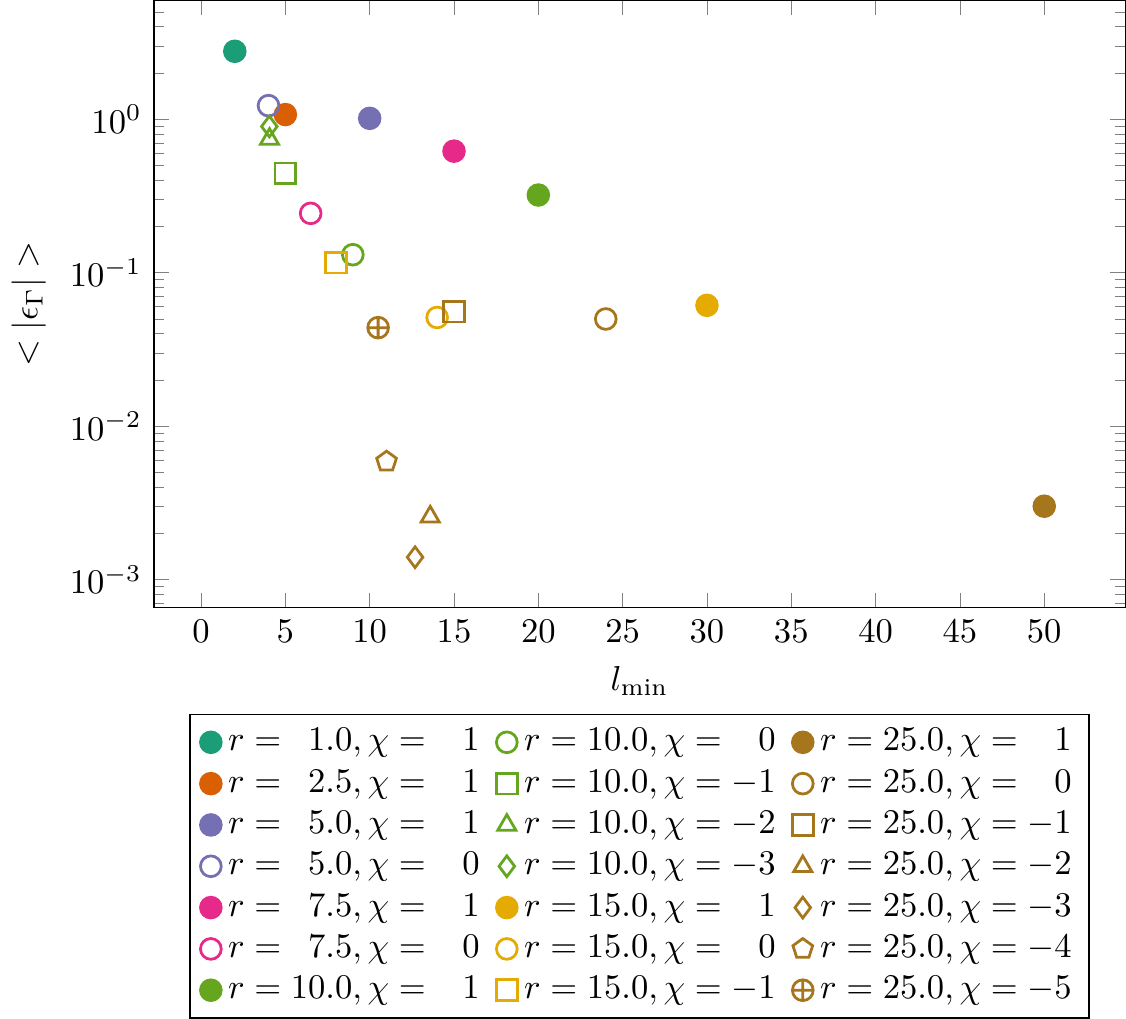}
\caption{Average absolute relative error $<\abs{\epsilon_{\Gamma}}>$ as a function
of the minimal characteristic length scale of the system, $l_{\txt{min}}$. In
the case of a pore without rods, this distance is twice the radius. When rods are
present within the pore this is the smallest distance between the pore wall and
a rod or between two different rods. Because the excess adsorption is a
derivative of the grand potential, the observed error is larger than
Figure~\ref{fig:omegaErrLJ}. 
\label{fig:gammaErrLJ}}
\end{figure}

Figure~\ref{fig:gammaErrLJ} shows the average absolute relative error
$<\abs{\epsilon_{\Gamma}}>$ as a function of the minimal characteristic length
scale of the system, $l_{\txt{min}}$. Because excess adsorption is a
derivative of the grand potential, the observed error is larger than
Figure~\ref{fig:omegaErrLJ}. Due to the more complex interactions in a
Lennard-Jones fluid, the error is also larger than the error observed for a
hard-sphere fluid in Figure~\ref{fig:gammaErrHS}. For pores without rods, it can
be observed that $\log{(<\abs{\epsilon_{\Gamma}}>)}$ scales almost linearly with
$l_{\txt{min}}$. Again, this scaling does not hold for pores with rods. This is
similar to what was observed for Figure~\ref{fig:omegaErrLJ}
which is probably at least partly caused by the limited contribution of topology
to the grand potential and the excess adsorption.

\begin{figure}
\center
\includegraphics[]{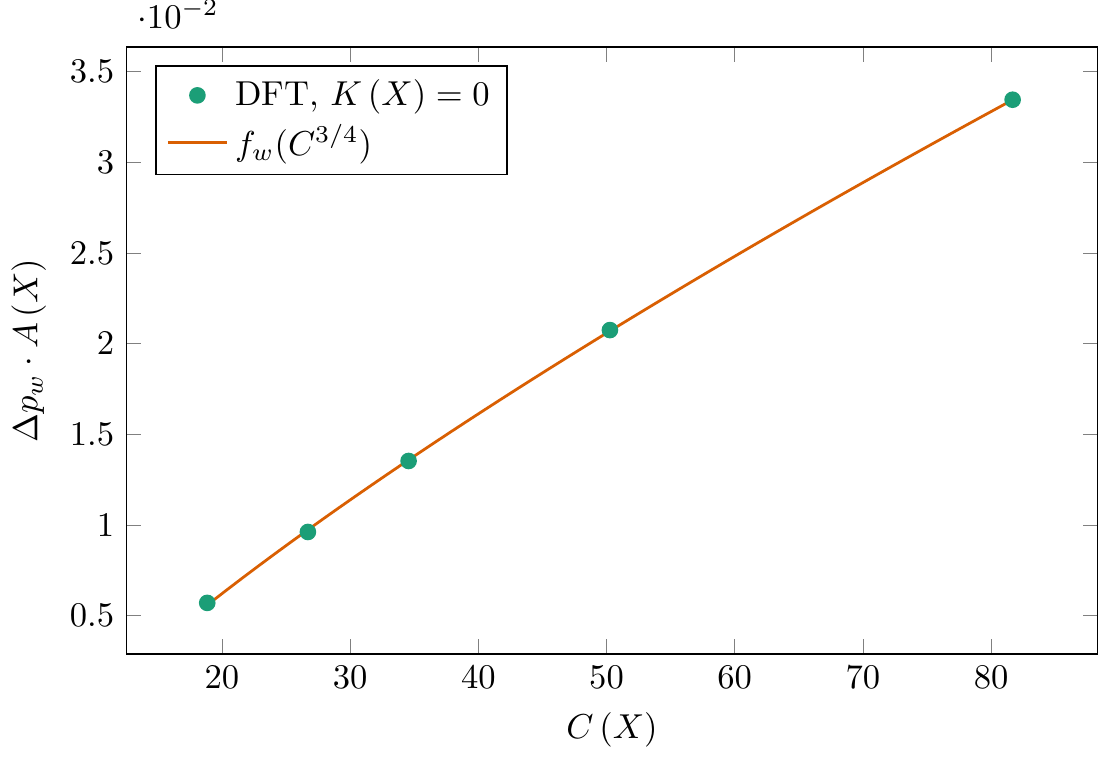}
\caption{Pressure difference between the second-order phase transition
associated with adsorption of gas onto a wall in the bulk and inside a pore
times the surface area Minkowski functional, $\Delta p_{w} \cdot A \brc{X}$, as
function of pore radius, $r_{p}$. The Minkowski functional for topology is zero
in all the shown simulations, $K \brc{X} = 0$.
The curve fit is equal to: $f_{w} (C^{3/4}) = \Omega_{\txt{lg}} +
\sigma_{\txt{lg}}' C^{3/4}$, where the Minkowski functional $C (X)$ only depends
on $r_{p}$.
The coefficients are: 
$\Omega_{\txt{lg}}  = -8.3   \cdot 10^{-3} \pm 0.1   \cdot 10^{-3}$ and 
$\sigma_{\txt{lg}}' =  1.538 \cdot 10^{-3} \pm 0.008 \cdot 10^{-3}$. 
\label{fig:wallRLJ}}
\end{figure}

As expressed in Equation~\ref{eqn:phase}, the Minkowski functionals can also be
used to evaluate the shift of the phase envelope as a function of morphology and
topology. Additionally, knowing the phase behavior also helps in
determining which terms should be used to expand the grand potential in terms of
the Minkowski functionals. Because the rod size is constant across different
simulations, the Minkowski functionals $A \brc{X}$ and $C \brc{X}$ can be
expressed as a functions of only the pore radius $r_{p}$ for constant bending
rigidity, $K \brc{X}$. This means that for $K \brc{X} = 0$, the pressure shift in
the phase envelope only depends on $r_{p}$. Figure~\ref{fig:wallRLJ} shows that
the pressure difference between the second-order phase transition associated with adsorption of gas
onto a wall in the bulk and inside a pore, $\Delta p_{w}$, can be predicted well
by the function: $f_{w} (C^{3/4}) = \Omega_{\txt{lg}} + \sigma_{\txt{lg}}'
C^{3/4}$, where the coefficients are: $\Omega_{\txt{lg}} = -8.3   \cdot
10^{-3} \pm 0.1   \cdot 10^{-3}$ and  $\sigma_{\txt{lg}}'    =  1.538 \cdot
10^{-3} \pm 0.008 \cdot 10^{-3}$. Although the power of $3/4$ is reasonably close to the
theoretical prediction of $1$, this shows that due to the small system size,
Hadwiger's additivity assumption starts to break down. A more accurate
fit of the grand potential can be found by adding an additional term
proportional to $C^{3/4}$ to the Minkowski functional expansion. This was done
for the results presented in Figures~\ref{fig:omegaErrLJ} -
\ref{fig:gammaErrLJ}. Another potential source of error could be that, for
very small pores, excluded volume effects keeping molecules away from the
wall could be significant.

\begin{figure}
\center
\includegraphics[]{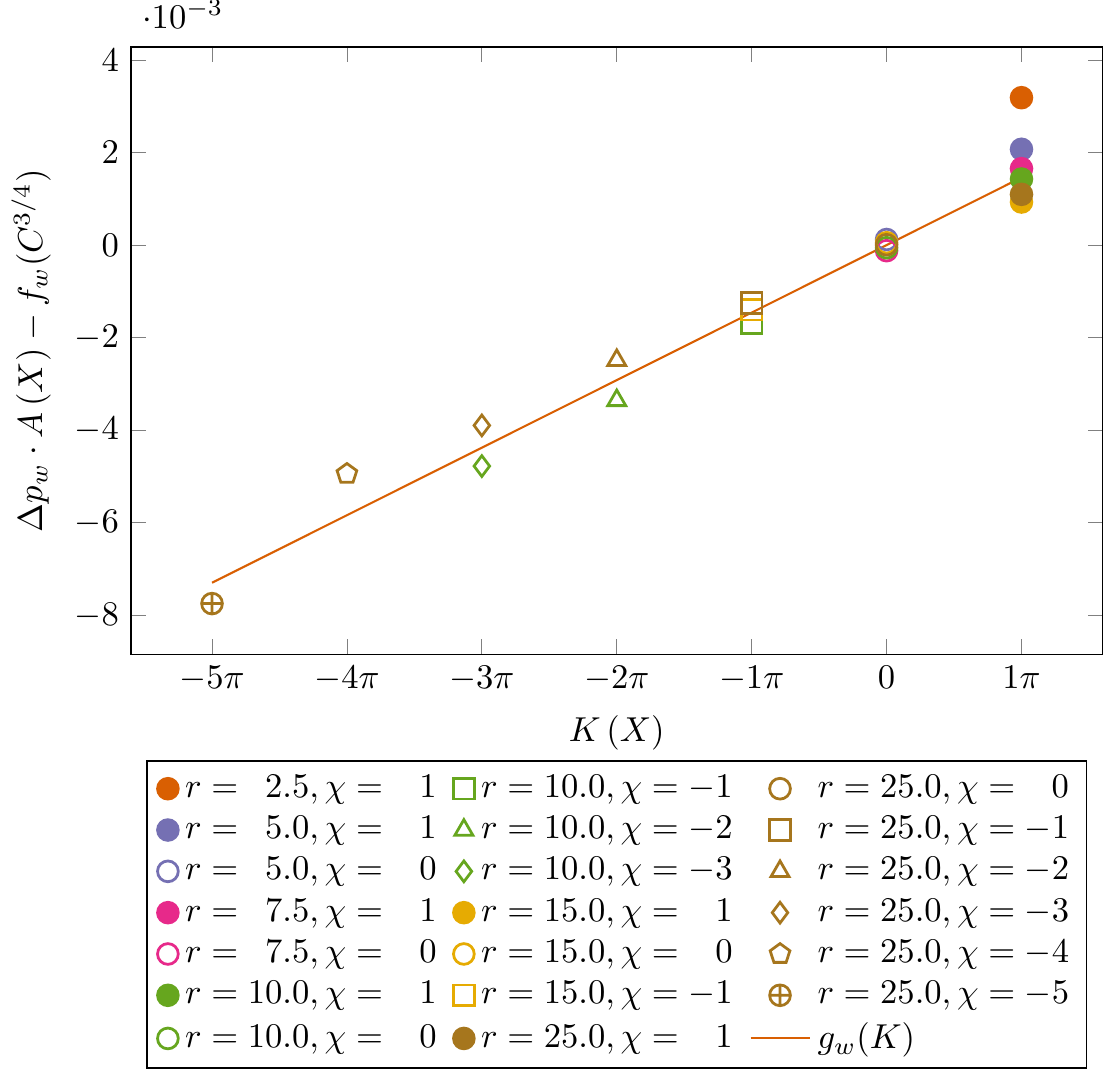}
\caption{Pressure difference between the second-order phase transition
associated with adsorption of gas onto a wall in the bulk and inside a pore
times the surface area Minkowski functional minus the function from
Figure~\ref{fig:wallRLJ}, $\Delta p_{w} \cdot A \brc{X} - f_{w} (C^{3/4})$, as
function of the Minkowski functional, $K \brc{X}$. The graph shows a collapse of
the data and a linear fit with: $g_{w} \brc{K} = \kappa_{\txt{lg}}' K \brc{X}$,
where $\kappa_{\txt{lg}}' = 4.6 \cdot 10^{-4} \pm 0.2 \cdot 10^{-4}$. 
\label{fig:wallKLJ}}
\end{figure}

To validate how well the function $f_{w} (C^{3/4})$ describes the data,
Figure~\ref{fig:wallKLJ} shows the pressure difference between the second-order
phase transition associated with adsorption of gas onto a wall in the bulk and
inside a pore times the surface area Minkowski functional minus the function
$f_{w} (C^{3/4})$, $\Delta p_{w} \cdot A \brc{X} - f_{w} (C^{3/4})$, as a function
of the Minkowski functional, $K \brc{X}$. The graph shows a collapse of the data
and, as expected, a linear fit with: $g_{w} \brc{K} =
\kappa_{\txt{lg}}' K \brc{X}$, where $\kappa_{\txt{lg}}' = 4.6 \cdot 10^{-4} \pm
0.2 \cdot 10^{-4}$. The collapse of the data confirms that the grand potential
and thus the second-order pressure shift in the phase envelope, $\Delta p_{w}$, is
proportional to $C^{3/4} \brc{X}$ and is linearly dependent on the topology of
the system $K \brc{X}$. In addition, the data shows that the sensitivity of the
grand potential to topology is about an order of magnitude smaller than the sensitivity to
the pseudo surface tension. The outlier, $r_{p} = 2.5$, $\chi = 1$,
confirms the breakdown of Hadwiger's theorem for small pore sizes. 

\begin{figure}
\center
\includegraphics[]{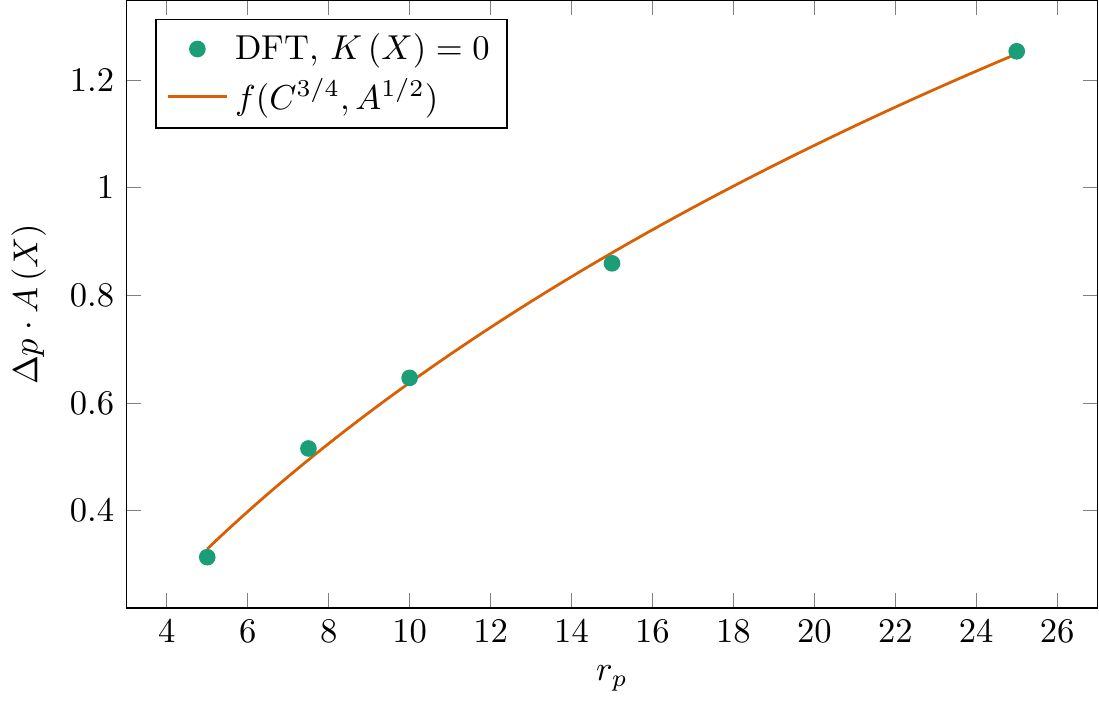}
\caption{Pressure difference between the capillary condensation pressure and bulk
phase transition pressure times the surface area Minkowski
functional, $\Delta p \cdot A \brc{X}$, as a function of pore radius, $r_{p}$. The Minkowski functional for
topology is zero in all the shown simulations, $K \brc{X} = 0$.
The curve fit is equal to: 
$f (C^{3/4},A^{1/2}) =  \Omega_{\txt{lg}} + \sigma_{\txt{lg}}' C^{3/4} +
p_{\txt{lg}}' A^{1/2}$, where the Minkowski functionals $C \brc{X}$ and $A \brc{X}$ only depend on $r_{p}$. The
coefficients are: $\Omega_{\txt{lg}} = -0.4 \pm 0.2$, $\sigma_{\txt{lg}}' = 0.12 \pm 0.03$, and
$p_{\txt{lg}}' = -0.04 \pm 0.02$.
\label{fig:phaseRLJ}}
\end{figure}

A similar analysis can be performed for the phase envelope shift of capillary
condensation, which is a first-order phase transition. Figure~\ref{fig:phaseRLJ}
shows the pressure difference between the capillary condensation pressure and
the bulk phase transition pressure times the surface area Minkowski
functional, $\Delta p \cdot A \brc{X}$, as a function of pore radius, $r_{p}$. The Minkowski functional for
topology is zero in all the DFT simulations shown in Figure~\ref{fig:phaseRLJ}, $K \brc{X} = 0$.
The curve fit is equal to: 
$f (C^{3/4},A^{1/2}) =  \Omega_{\txt{lg}} + \sigma_{\txt{lg}}' C^{3/4} +
p_{\txt{lg}}' A^{1/2}$, where the Minkowski functionals $C \brc{X}$ and $A \brc{X}$ only depend on $r_{p}$. The
coefficients are: $\Omega_{\txt{lg}} = -0.4 \pm 0.2$, $\sigma_{\txt{lg}}' = 0.12 \pm 0.03$, and
$p_{\txt{lg}}' = -0.04 \pm 0.02$.
The curvature of the simulation data presents a clear case for adding an additional
term proportional to $A^{1/2} \brc{X}$ to the expansion of the grand potential with Minkowski functionals.

\begin{figure}
\center
\includegraphics[]{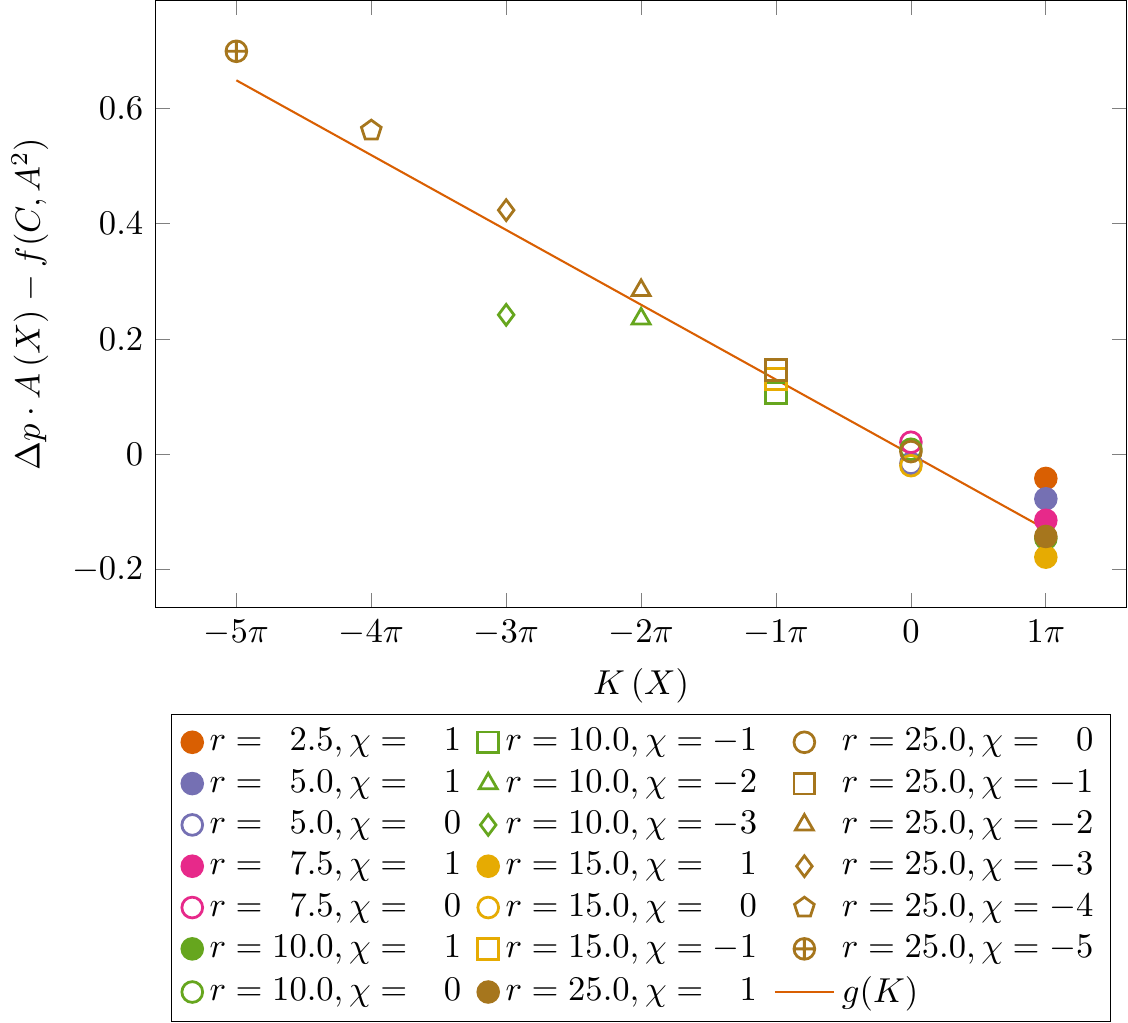}
\caption{Pressure difference between capillary condensation pressure and bulk
phase transition pressure times the surface area Minkowski
functional minus the function from Figure~\ref{fig:phaseRLJ}, $\Delta p \cdot A
\brc{X} - f (C^{3/4},A^{1/2})$, as a function of the Minkowski functional, $K \brc{X}$. The graph shows a
collapse of the data and a linear fit with: $g (K) = \kappa_{\txt{lg}}' K \brc{X}$,
where $\kappa_{\txt{lg}}' = -0.041 \pm 0.002$. 
\label{fig:phaseKLJ}}
\end{figure}

To validate the data fit shown in Figure~\ref{fig:phaseRLJ}, Figure~\ref{fig:phaseKLJ}
shows the pressure difference between capillary condensation pressure and the bulk
phase transition pressure times the surface area Minkowski
functional minus the function from Figure~\ref{fig:phaseRLJ}, $\Delta p \cdot A
\brc{X} - f (C^{3/4},A^{1/2})$, as a function of the Minkowski functional, $K \brc{X}$.
The graph shows a collapse of the data and, as predicted by theory, a linear fit
with: $g_{w} \brc{K} = \kappa_{\txt{lg}}' K \brc{X}$, where $\kappa_{\txt{lg}}'
= 4.6 \cdot 10^{-4} \pm 0.2 \cdot 10^{-4}$. The collapse of the data into one
single line, confirms that the grand potential is proportional to $C^{3/4} \brc{X}$, $A^{1/2}
\brc{X}$, and $K \brc{X}$. The fitting parameters $\sigma_{\txt{lg}}'$ and
$p_{\txt{lg}}'$ show that the difference between the capillary condensation
pressure and the bulk phase transition pressure is the most sensitive to changes
in the pseudo surface tension and the pseudo pressure. The  sensitivity to
topology changes, represented by the parameter $\kappa_{\txt{lg}}'$, is much
smaller. The outlier, $r = 10.0$, $\chi = -3$, is a system where the topology
change compared to, $\chi = 1$, caused an additional phase transition to occur.
Surprisingly, a difference between Figure~\ref{fig:wallKLJ}
and~\ref{fig:phaseKLJ} is that the phase envelope shift dependence on topology
is different depending on whether the phase transition is a first-order or a
second-order phase transition. This is a topic for further research.

\section{Conclusions}

We studied the effect of morphology and topology on capillary
condensation in a systematic manner using a Minkowski functional framework.
Consistent with the literature \cite{konig2004}, it is found that hard-sphere
fluids obey Hadwiger's theorem \cite{hadwiger1957} down to quite small pore
sizes, especially when using the Minkowski functionals to reconstruct the grand
potential. For the excess adsorption, a significantly larger average absolute
error between the DFT simulations and the Minkowski functional reconstruction of
the excess adsorption is found.

Analyzing the error for various geometries, it is observed that the error decreases
rapidly with increasing pore size. However, it is also found that it is not trivial to
find a characteristic length scale of the system at which Hadwiger's theorem
starts to break down. While increasing the size of a rod inside a pore has a
clear effect on the error, moving a rod around inside a pore while keeping the
Minkowski functionals of the system constant has no effect. This can potentially
be explained by the fact that topology only has a small contribution to the
grand potential for the systems studied in this work. 

Changing the system from a hard-sphere fluid to a Lennard-Jones fluid, the
interaction length becomes much longer and phase behavior becomes much more
complex. This results in larger errors in both the grand potential and the
excess adsorption, which can be partly overcome by introducing additional terms
in the expansion of the grand potential. Trends in the error observed
for hard-sphere fluids are also observed for Lennard-Jones fluids.

Analyzing the contributions of the various Minkowski functional coefficients to
the grand potential shows that wall adsorption is dominated by a surface-tension
term and pseudo pressure, which is proportional to $\propto A^{1/2}$. During
capillary condensation, a pseudo surface tension term proportional to $\propto
C^{3/4}$ also gains in importance and this regime is characterized by large
discontinuities in the coefficients which match the discontinuities caused by
capillary condensation. After the bulk phase transition the system is
increasingly dominated by the pressure. A similar analysis of the coefficients
contributing to the excess adsorption shows the pseudo pressure as a positively
contributing term to the wall adsorption and the surface tension as a negatively
contributing term. The reason that there are both positive and negative
contributions to the excess adsorption is due to the fact that the derivatives
present in the Minkowski functional representation of the excess adsorption can
be both positive and negative. During capillary condensation, the pseudo surface
tension term becomes significantly more important. For both the grand potential
and the excess adsorption the effect of topology is modest.

Last but not least, the effect of confinement on phase behavior is investigated. It is found
that pressure shift in the second-order phase transition describing adsorption
on the pore wall is proportional to $C^{3/4} \brc{X}$ and $K \brc{X}$, which is
close to what is predicted based on Hadwiger's theorem. However, the pressure
for the first-order phase transition describing capillary condensation is
proportional to $C^{3/4} \brc{X}$, $C^{1/2} \brc{X}$ and $K \brc{X}$, which is a
significant deviation from theory. In addition, the first-order phase transition
and the second-order phase transition have the opposite dependence on
topology. For first-order phase transitions as a function of the Minkowski
functional, $K \brc{X}$, the fitting parameter is positive, while for
second-order phase transitions this parameter is positive. Whether this is a
finding which holds in general for first- and second-order phase transitions has
to be investigated further.

The Minkowski functionals provide a useful framework to study capillary
condensation. The separation of geometry and thermodynamics allows for a method
to systematically study the effect of surface area, circumference, and the Euler
characteristic on phase behavior. This provides many opportunities for
future research. One of the many open questions is whether the different effect
that topology has on first- versus second-order phase transitions is also found
for different Lennard-Jones fluids, or even for first- and second-order phase
transitions in general. Another question is whether it is possible to use the
Minkowski functionals for upscaling. The idea is to perform a number of
simulations on different small geometries with known Minkowski functionals to compute the
Minkowski functional coefficients for the excess adsorption. These coefficients
are then used to predict the excess adsorption for a much larger experimental
disordered porous medium with known Minkowski functionals. Other questions to
consider working on are: to continue looking for a characteristic length scale
that describes when Hadwiger's theorem breaks down, study whether sorption
hysteresis can be described using the Minkowski functionals, and to study
whether the Minkowski functionals can be used for higher molecular weight
molecules.

\section{Acknowledgements}

This work was supported as part of the Center for Mechanistic Control of
Water-Hydrocarbon-Rock Interactions in Unconventional and Tight Oil Formations
(CMC-UF), an Energy Frontier Research Center funded by the U.S. Department of
Energy, Office of Science under DOE (BES) Award DE-SC0019165.

The majority of the computing for this project was performed on the Mazama
cluster of the Center for Computational Earth \& Environmental Science (CEES) at
Stanford University. We would like to thank Stanford University and the Stanford
Research Computing Center for providing computational resources and support that
contributed to these research results.

\section*{References}

\bibliographystyle{unsrt}

\end{document}